\documentclass[preprint]{aastex}

\usepackage{epsfig}
\usepackage{booktabs}
\def\micron {$\mu$m}
\def \  {$\dagger$}

\def \ch {\checkmark}
\def \dch {\checkmark\checkmark}

\begin{document}

\title{Absorption Efficiencies of Forsterite. I:  DDA Explorations in Grain Shape and Size}
\author{Sean S. Lindsay$^1$, Diane H. Wooden$^2$, David E. Harker$^3$, Michael S. Kelley$^4$, Charles E. Woodward$^5$, Jim R. Murphy$^6$}
\altaffiltext{1}{Department of Earth and Planetary Sciences, University of Tennessee, 1421 Circle Drive, Knoxville, TN 37996-2366; slindsay@utk.edu}
\altaffiltext{2}{Space Science Division, NASA Ames Research Center, MS 245-3, Moffett Field, CA 94035-0001, USA; diane.h.wooden@nasa.gov}
\altaffiltext{3}{Center for Astrophysics and Space Sciences, University of California, San Diego, 9500 Gilman Drive, La Jolla, CA, 92093-0424, USA; dharker@uscd.edu}
\altaffiltext{4}{Department of Astronomy, University of Maryland, College Park, MD 20742, USA; msk@astro.umd.edu}
\altaffiltext{5}{Minnesota Institute of Astrophysics, 116 Church Street, S. E., University of Minnesota, Minneapolis, MN 55455, USA; chelsea@astro.umn.edu}
\altaffiltext{6}{Department of Astronomy, New Mexico State University, P.O. Box 30001, MSC 4500, Las Cruces, NM 88003-8001; seanslindsay@gmail.com}

\begin{abstract}
We compute the absorption efficiency ($Q_{\rm abs}$) of forsterite using the discrete dipole approximation (DDA) in order to identify and describe what characteristics of crystal grain shape and size are important to the shape, peak location, and relative strength of spectral features in the 8~--~40~\micron \ wavelength range.  Using the  DDSCAT code, we compute $Q_{\rm abs}$ for non-spherical polyhedral grain shapes with $a_{\rm eff}$ = 0.1~\micron.  The shape characteristics identified are: 1) elongation/reduction along one of three crystallographic axes;  2) asymmetry, such that all three crystallographic axes are of different lengths; and 3) the presence of crystalline faces that are not parallel to a specific crystallographic axis, e.g., non-rectangular prisms and (di)pyramids. Elongation/reduction dominates the locations and shapes of spectral features near 10, 11, 16, 23.5, 27, and 33.5~\micron, while asymmetry and tips are secondary shape effects.  Increasing grain sizes (0.1~--~1.0~\micron) shifts the 10, 11~\micron \ features systematically towards longer wavelengths and relative to the 11~\micron \ feature increases the strengths and slightly broadens the longer wavelength features.  Seven spectral shape classes are established for crystallographic a-, b-, and c-axes and include columnar and platelet shapes plus non-elongated or equant grain shapes.  The spectral shape classes and the effects of grain size have practical application in identifying or excluding columnar, platelet or equant forsterite grain shapes in astrophysical environs.   Identification of the shape characteristics of forsterite from 8~--~40~\micron \ spectra provides a potential means to probe the temperatures at which forsterite formed.  
\end{abstract}
\keywords{comets: general -- circumstellar matter --  infrared: planetary systems -- lines: profiles -- methods: numerical -- protoplanetary disks}

%%%%%%%%%%%%%%%%%%%%%%%%%%%%%%%%%%%%
\section{INTRODUCTION}
\label{sec:intro}

The silicate crystal mass fraction, $f_{\rm cryst}$, which is the mass ratio of silicate crystals to all silicates, is an analytical tool that can be used to study the taxonomy of comets and the evolution and radial stratification of proto-planetary disks. Crystalline silicates are rare in the ISM, $\lesssim$ 2.2\% along the line-of-sight to the Galactic Center and $\lesssim$ 5\% along other lines-of-sight \citep{Li:2001}.  In contrast, primitive Solar System materials reveal significantly higher $f_{\rm cryst}$.  Thermal models of comet comae produce values of $f_{cryst}$ as high as $\approx 0.8$ \citep{Harker:2002,Harker:2004err} and generally comets have  $f_{\rm cryst} \gtrsim 0.2$ \citep{Kelley:2009}.  \emph{Stardust} samples of comet 81P/Wild~2 include large ($\sim$ 20 \micron) single mineral crystals \citep{Brownlee:2006} and  X-ray microprobe analysis indicates a $f_{\rm cryst} > 0.5$ \citep{Ogliore:2009}.   These crystals are thought to be condensates \citep{Brownlee:2006,Nakamura:2008,Tsuchiyama:2009}.  The discrepancy between ISM and Solar System abundances of crystalline silicates indicates the Solar System crystals must be a product of the formation and evolution of a dusty disk with a radial thermal gradient.  The hot inner regions of proto-planetary disks have temperatures (T $\gtrsim 1000-1400$~K) appropriate for rapid, direct gas phase condensation or annealing of amorphous ISM silicates into crystalline silicates \citep{Bell:2000,Gail:2002,Gail:Astromineralogy,Henning:2003Astromineralogy,Wooden:2007PPV,Henning:2010ARAA}.  The presence of these hot refractory materials in cold, thermally unaltered comets indicates an efficient process of transporting materials from the hot inner regions of the disk to the cold outer regions \citep{Ciesla:2007,Hughes:2010trans}.  Further, the abundance of crystalline silicates at different heliocentric distances serves as a tracer of the efficiency of radial transport. Hence, $f_{\rm cryst}$ provides constraints for radial transport in protoplanetary disk models \citep{Wooden:2007PPV,Wooden:2008}.  In turn, radial transport in disks is used to interpret where, how, and at what time(s) and over what duration(s) the crystals formed and over what subsequent timescales they were transported outward into the comet forming zones of the Solar proto-planetary disk \citep{Gail:2004, Ciesla:2010, Hughes:2010trans}.

The increasing sophistication of remote sensing observational techniques and advances in the analytic interpretation of infrared (IR) spectral energy distributions (SEDs) emitted by dust in comet comae, as well as constraints imposed by collected materials in comet sample-return missions, now necessitate modernization of astrophysical dust grain models to accurately derive $f_{\rm cryst}$.  As the signal-to-noise quality of observations has improved over the years through space-based facilities such as the \emph{Spitzer Space Telescope}, new subtleties in the IR silicate emission features evident in SEDs for cometary comae are discernible.  Current modeling techniques are insufficient and cannot replicate many of the complex spectral signatures in the observed SEDs.  Recent laboratory and modeling efforts suggest that to account for crystalline resonances, it does not suffice to only consider grain composition, but it is required to also consider grain shape and size.  Grain shape and size can change the wavelength location, shape, and relative strength of the crystalline silicate features present from 8~--~40 \micron \ \citep{Fabian:2001, Min:2005HB,Anderson:2006, Tamanai:2006ApJ, Tamanai:2006JQSRT, Imai:2009, Mutschke:2009, Tamanai:2009CDNF, Koike:2010}.  The magnitude of the changes in peak wavelength location due to the variation of grain shape in these laboratory experiments is on the order of the changes in peak location arising from the addition of up to 25\% iron to an Mg-rich olivine crystal \citep{Koike:2003,Kelley:2009}.  Thus, the grain shape of the crystals must be considered in parallel to external grain composition when modeling the dust in comet comae, especially when the derivation of $f_{\rm cryst}$ is sought.   The shape of the crystals in comet dust could be indicative of the process of crystal formation either by gas phase condensation \citep{Tsuchiyama:1998min,Bradley:1983,Kobatake:2008}, selective evaporation \citep{Tsuchiyama:1998,Takigawa:2009}, or annealing \citep{Koike:2006,Koike:2010}.  Furthermore, if variations in $f_{\rm cryst}$ are used to interpret radial transport efficiencies or dust evolution processes in proto-planetary disks (PPDs) (e.g., \citealp{Manoj:2011,Hughes:2010trans}), accurate model descriptions of the dust, including the crystalline component, are required \citep{Olofsson:2009,Oliveira:2011}.

In this paper, we discuss results of new computational modeling of polyhedral shaped forsterite (Mg$_2$SiO$_4$, the magnesium-rich end-member of the olivine group) using the discrete dipole approximation (DDA). We investigate the characteristics of grain shape and grain size that produce subtleties in synthetic thermal IR SEDs. Such crystalline silicate features are now routinely observed by remote sensing techniques in comets and proto-planeteary disks conducted at moderate spectral resolutions ($R \equiv \Delta\lambda/\lambda \gtrsim$ 1000) and high sensitivity (continuum flux densities of $mJy$ to $\mu Jy$).  We present the initial stages of building a bottom-up model that will compute absorption efficiencies for large, porous, multimineralic aggregates with crystalline inclusions for the 8~--~40~\micron \ IR spectral range.  We systematically vary the axial ratios and external shapes to enable an in depth study of what characteristics of grain shape are important to the crystalline resonant features in the absorption efficiency spectra for forsterite, which is a dominant mineral in comet dust \citep{Harker:2007,Lisse:2007,Reach:2010,Harker:2011}, which produces narrow-band crystalline resonance spectral features in the 8~--~40 \micron \ region.  The crystal shapes selected from this study provide realistic model crystals that can be incorporated in the future into a large, porous, multimineralic aggregates used in thermal dust models.

A major challenge in deriving the silicate mineralogy of comets is understanding how the anisotropic nature of forsterite crystals affects the 8 - 40~\micron \ spectral features' shape, strength, and peak position.  As forsterite is the main crystalline silicate identified in cometary comae \citep{Campins:1989, Crovisier:1997, Wooden:1999,Harker:2002,Harker:2004err, Lisse:2006Sci,Lisse:2007}, accurate models for forsterite's emission/absorption features are the primary requirement to determine the silicate mineralogy of comets.  However, there exists a dearth of models that fully account for the anisotropic nature of forsterite.  Here we present a new modeling investigation into the 8 - 40 \micron \ absorption resonance features of forsterite using the discrete dipole approximation (DDA) code DDSCAT \citep{Draine:1994DDA,Draine:2008}, which fully accounts for the anisotropic nature of forsterite.  

%HERE - dec. 9
The paper proceeds in the following manner. In \S~\ref{sec:comp_approach}, we briefly summarize three commonly used computational approaches to modeling grains and explain why DDA is the method chosen to compute polyhedral shaped forsterite crystals.  In \S~\ref{sec:meth}, we describe the details of how DDA is applied to our study of grain shape characteristics, and what constraints apply to using the publicly available DDA code DDSCAT \citep{Draine:1994DDA,Draine:2008}.  Section~\ref{sec:results} presents the DDSCAT-calculated absorption efficiencies ($Q_{\rm abs}$).  We compute $Q_{\rm abs}$ for three grains shape characteristic exercises: 1) the elongation/reduction of a single crystallographic axis; 2) rectangular prism shapes that have three different side lengths, and are therefore asymmetric; and 3) shapes that have different external geometries but have the same crystallographic axis ratio with an emphasis on exploring the effects of (di)pyramidal structures through a comparison of $Q_{\rm abs}$ between prism~--~dipyramidal pairs.  Also in \S~\ref{sec:results}, we compute $Q_{\rm abs}$ for variations in grain size from 0.1~--~3.0 \micron \  in effective radius.  In \S~\ref{sec:disc}, we expand upon the results of \S~\ref{sec:results} and discuss the implications of crystal shape characteristics and their potential to act as a probe of formation enivrons, and we provide a practical application to the results to the case of cometary comae.  In \S~\ref{sec:summary} we summarize the major conclusions.  The major forsterite bands we investigate are those located near 10, 10.5, 11, 16, 19, 23, 27, and 33 \micron.  In addition, we examine more subtle emission features appearing at approximately 11.9 and 25.2 \micron.

%%%%%%%%%%%%%%%%%%%%%%%%%%%%%%%%%%%%%%%%%%
\section{COMPUTATIONAL APPROACHES TO OPTICAL CONSTANTS}
\label{sec:comp_approach}

There are three broad approaches in modeling methods to compute the optical properties arising from an ensemble of small (less than  a few microns in size) discrete grains that are subsequently incorporated into thermal grain models to replicate observed SEDs \citep[for a review see][]{Min:2009CDNF}.  The first approach is to assume that the grains are homogeneous in composition and spherical in shape.  In this case, Mie theory \citep{Mie:1908} can be used, and computations are simple and fast.  However, in cometary samples from interplanetary dust particles \citep[][e.g., ]{Molster:2002iii,Bradley:2003} 
crystalline grains are non-spherical.  Forsterite grains in \emph{Stardust} samples can have euhedral, polyhedral with well-defined sharp faces and edges, shapes \citep{Nakamura:2012} or irregular shapes (i.e., not even or balanced in shape or arrangement) where irregular shapes may arise due to capture in aerogel \citep{Zolensky:2008}.  \citet{Yana:1999} and \citet{Fabian:2001} demonstrate that Mie computations using spherical particles inadequately replicate spectrophotometric observations if crystalline silicates are present within the emitting dust ensemble.  The symmetry of a sphere creates resonance effects at particular values of the index of refraction that are not seen in natural grains with irregular shape \citep{Yana:1999,Fabian:2001,Min:2009CDNF,Henning:2010}.

The second modeling methodology is a statistical approach to account for irregular, non-polyhedral, grain shapes.  In this approach, an arbitrarily shaped particle is represented by a distribution of simple particles that have exact solutions. For example, the continuous distribution of ellipsoids (CDE) method  \citep{Fabian:2001,Min:2003,Bohren:1983} is frequently used.  This adopts a statistical distribution of ellipsoids, which may include spheres, that can vary in size (within the Rayleigh limit), principle axes ratio (or for crystals, crystallographic or optical axes), and orientation.  The absorption characteristics for each ellipsoid can be exactly computed, and the grain shape irregularity is defined by the choice of ellipsoids.   Within the statistical approach, ellipsoids are but one of several simple shapes with an exact solution that can be chosen.  Other notable simple shapes that can be used to acquire the absorption characteristics of an irregularly shaped grain are hollow spheres \citep{Min:2003,Min:2005AA} and gaussian random spheres \citep{Muinonen:1996, Volten:2001,Mutschke:2009}.  

The third modeling methodology seeks to directly calculate the optical properties for arbitrarily shaped grains.  These computationally intensive approaches are the DDA \citep{Purcell:1973DDA} method, the coupled dipole approximation (CDA) \citep{Min:2008}, or the T-Matrix method (TMM) \citep{Mishchenko:1996}.  The TMM \citep{Mishchenko:2007reflist} expands the incident and scattered electric fields in series of suitable vector spherical wave functions, wherein the relation between the columns of the respective expansion coefficients is established by means of a transition matrix (or T-matrix). This approach can be applied to the entire scatterer as well as to separate parts of a composite scatterer \citep{Mishchenko:2007reflist}.  The TMM has been successful in modeling the polarization of light scattered by aggregate grains \citep{Kolokolova:2007, Kolokolova:2010}.  In theory the TMM can account for anisotropic materials \citep{Schmidt:2009}, however, to date the TMM has not been extended  to model a discrete tri-axial forsterite crystal due to a reliance upon a reformulation of the TMM equations that are specific to a particular grain shape.

The DDA \citep{Purcell:1973DDA, Draine:1994DDA, Min:2006JQSRT} is an absorption and scattering modeling method that allows for grains of arbitrary shape and crystalline composition such that all three crystallographic axes are accounted for simultaneously.  With DDA the user can specify the geometry of the target (model grain), the size of the target, the alignment of multiple indices of refraction with respect to the target geometry, the wavelength range over which the model is to run, and the composition of individual dipoles.  With this high level of control, realistic targets of discrete crystalline grains can be constructed, permitting calculation of the target's absorption and scattering efficiencies. 

Our goal in this paper is to identify and describe characteristics of grain shape that affect the absorption of forsterite crystals in the context of forsterite's anisotropic shape (i.e., three distinct indices of refraction associated with three crystallographic axes).  This necessitates a careful control on the exact shape of the target and a method that allows for computations to be done incorporating all three crystallographic axes simultaneously.  The DDA approach satisfies both of these needs and is therefore selected for our study.

%%%%%%%%%%%%%%%%%%%%%%%%%%%%%%%%%%%%%
\section{METHODOLOGIES}  
\label{sec:meth}

The DDA technique is used to model crystalline forsterite (Mg$_2$SiO$_4$) between 8~--~40 \micron \ with a wavelength spacing of $\Delta\lambda$ = 0.05 \micron.  The DDA method represents an arbitrarily shaped particle by a collection of dipoles arranged on a three-dimensional lattice.  In essence, each dipole is a polarizable element that is a coarse representation of a large number of actual atoms within the target, such that with the appropriate indices of refraction the bulk dielectric constant of the material is reproduced by the dipoles \citep{Purcell:1973DDA}. Each dipole reacts to incident radiation producing a dipole field that in turn then interacts with all other dipoles in the DDA lattice or grid of dipoles that represents of the grain.  This method is computationally intensive, yet also the most accurate for calculating absorption and scattering properties of grains. We use the publicly available DDSCAT code \citep[Version 7.0]{Draine:2008_70}, which has been parallelized to run on the NASA Advanced Supercomputing (\emph{NAS}) facility, \emph{Pleiades}.\\

%%%%%%%%%%%%%%%%%%%%%%%%%%%%%%%%%%%%%
\subsection{DDA Targets} 

Since the user specifies the location of each individual dipole in DDSCAT, any target shape can be created.  This allows us to create a variety of target shapes that can be used to explore three main shape characteristics: 1) the elongation or reduction of a crystal along a single crystallographic axis; 2) shapes that differ in length for all three crystallographic axes,  (i.e., crystallographic axial ratio), or rather are tri-axially asymmetric; and 3) the presence of faces and edges that are not aligned with a single crystallographic axis.  The targets used to explore these shape characteristics are polyhedral in shape and include simple prisms and more complex dipyramidal structures that can be combinations of prismatic shapes with pyramidal tips. 

In addition to user-specified dipole location, any index of refraction with up to three differing crystallographic axes can be applied on a dipole-by-dipole basis, allowing for a unique composition to be applied to each dipole.  Forsterite is anisotropic such that it has three distinct crystallographic axes, $\alpha,~\beta,~\gamma$, commonly referred to as the a-, b-, and c-axes.  When creating our DDA targets, we are able to assign each dipole three indices of refraction that are each associated with one of forsterite's crystallographic axes.  The crystallographic axes in the targets are assigned such that they are parallel to the x-, y-, and z- geometric axes.  Using a variety of assignments, we are able to create targets that are elongated or reduced along our choice of crystallographic axis.  These assignments allow for the investigations into the spectral effects associated with precise crystallographic axis ratios.

\begin{figure}[!p]       %Fig. 1
	\begin{center}
		\epsfig{file=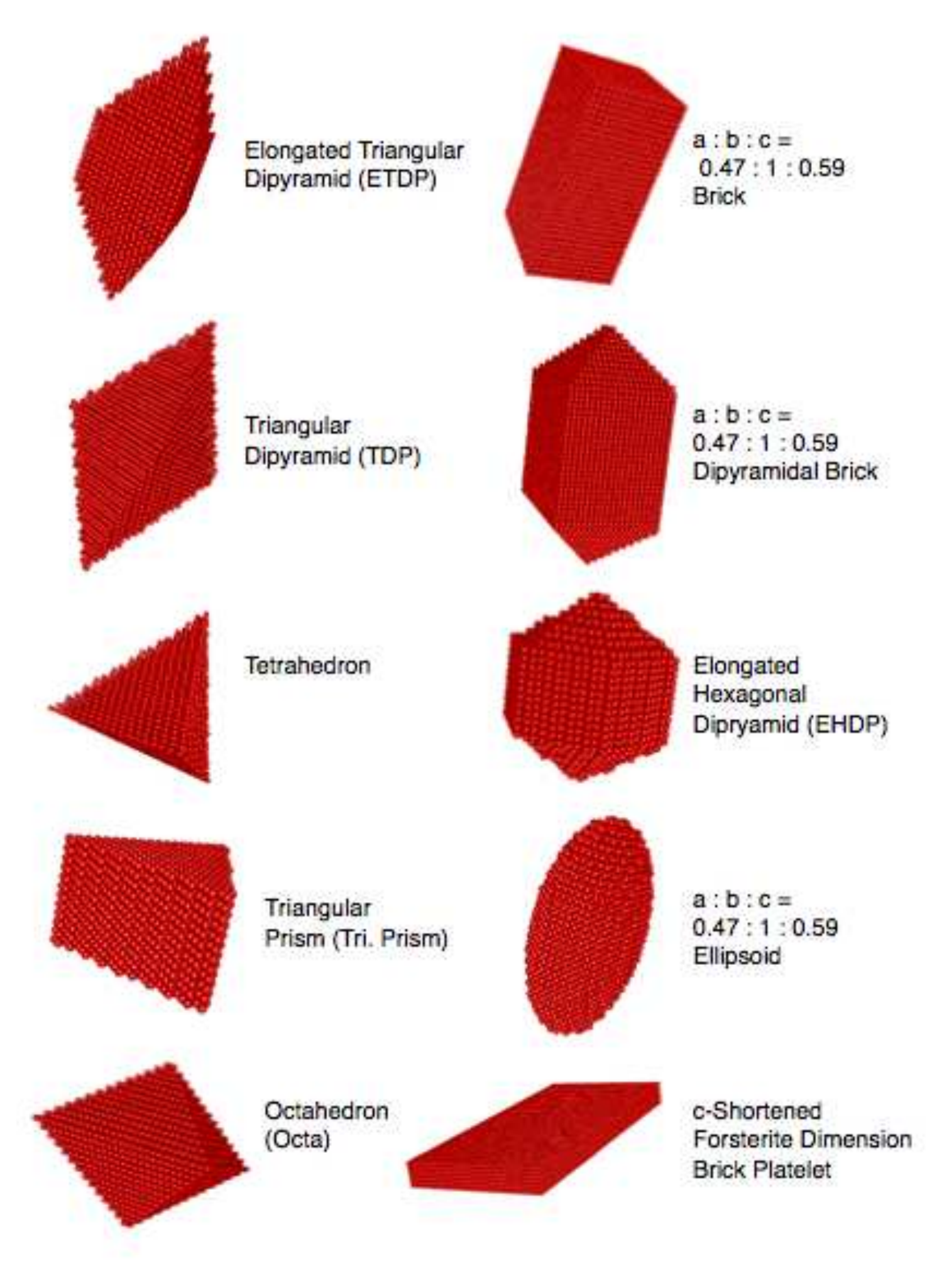,height=7.5in,width=6.5in}
		\caption{A selection of some of the target shapes used in this study.  Each sphere indicates the placement of a dipole in DDSCAT.  The listed axial ratios for the ellipsoid, brick, and dipyramidal brick are a~:~b~:~c $=$ 0.47~:~1~:~0.59, which is the axial ratio for the unit cell of forsterite.}
	\label{fig:target_shapes_collection}
	\end{center}
\end{figure}

A selection of the types of DDA targets used in this study are provided in Fig.~\ref{fig:target_shapes_collection}.  The entire sample of polyhedral shapes  includes shapes elongated along a single crystallographic axis (columnar), reduced along a single crystallographic axis (platelet), with all three axes unequal in length (asymmetric),  prismatic shapes, (di)pryamidal shapes, and elongated dipyramidal shapes.

%%%%%%%%%%%%%%%%%%%%%%%%%%%%%%%%%%%%%
\subsubsection{Numerical Constraints}  

There are two main numerical constraints that apply to the DDSCAT code.  The first is that the lattice spacing $d$ between dipoles must be small enough, and the number of dipoles used large enough, to accurately describe the shape of the target.  This constraint is key for this work because we use a variety of polyhedral grain shapes.  As such, we use a sufficiently large number of dipoles so that the target has relatively smooth faces and well-defined edges.  Each dipole in the lattice is represented by a cubical volume element with side length, $d$.  Thus, a sufficient number of these cubical volume elements are used to ensure that each cubical volume element is small compared to the overall size and shape of the target.  The second constraint is stringent, and dictates that in order for the DDA to apply, $d$ must be small compared to the wavelength of the radiant energy passing through the specified medium characterized by a complex index of refraction, $m$.  This is satisfied by the condition $|m|kd < 1.0$ for $Q_{\rm abs}$ calculations, where $k = 2\pi/\lambda$ is the wavenumber \citep{Draine:1994DDA}.  If scattering calculations are needed, the condition is tighter, such that $|m|kd < 0.5$. 

A more useful form of the latter constraint relates $|m|kd < 1.0$ to the number of dipoles that are required to sufficiently model a target grain of a certain grain size.  DDSCAT models a target on a three--dimensional lattice, where each cubic volume element with side length $d$ is assigned to be a dipole with a particular index of refraction.  Thus, each dipole element has volume $d^3$ and the total target volume is $V = N_{\rm dip}d^3$, where $N_{\rm dip}$ is the total number of dipoles in the target.  This volume must be equivalent to the actual physical volume of the polyhedral target.  If the volume of the polyhedral shaped target were collapsed into a sphere of equivalent volume, then the radius of that sphere defines the \emph{effective radius} of the target, $a_{\rm eff}$, so that $a_{\rm eff} = d~[(3N_{\rm dip})/(4\pi)] ^{1/3}$.  Expressing $a_{\rm eff}$ in terms of $N_{\rm dip}$ and $|m|kd$, yields

\begin{center}
\begin{equation}
	a_{\rm eff} = \left(\frac{3N_{\rm dip}}{32\pi^4}\right)^{\frac{1}{3}}\frac{\lambda}{|m|}|m|kd
\end{equation}
\end{center}

\noindent where $\lambda$ is the wavelength of radiant energy being considered.  Applying the requirement for accurate absorption efficiency calculations, $|m|kd < 1.0$, gives us a relation for the number of dipoles required for a target as a function of the incident radiation's wavelength, the complex index of refraction for the target material, and for the grain effective radius:

\begin{center}
\begin{equation}
	N_{\rm dip} > \frac{32\pi^4}{3}\left(\frac{|m|}{\lambda}a_{\rm eff}\right)^3
\end{equation}
\end{center}

\noindent Computing $Q_{\rm abs}$ for a minimum wavelength of 3~\micron \ and assuming an index of refraction for forsterite at 3 \micron \ of 1.8851 for the real part of the index of refraction \citep{Steyer:1974thesis} implies a target with an effective radius of 1~\micron \ requires 258 dipoles;  2~\micron \ requires 2,063 dipoles; 3~\micron \ requires 6,961 dipoles; and 5~\micron \ requires 32,224 dipoles.  For grain radii $a_{\rm eff} \lesssim 3.0$~\micron, we utilize roughly 2,000~--~10,000 dipoles, depending on the target, for the full 3 -- 40~\micron \ wavelength range.

%%%%%%%%%%%%%%%%%%%%%%%%%%%%%%%%%%%%%%%%%%%%%%%%%%
\subsubsection{Forsterite Spectral Resonances}  

We seek to delineate how varying the shape and size characteristics of forsterite crystals affects the spectral crystalline resonant features' shape, peak location, and relative strength.  By Kirchhoff's law of thermal radiation, the spectral features present in the absorption efficiencies ($Q_{\rm abs}$) will be present as emission features when these same grains are in radiative equilibrium with sunlight in comet comae \citep{Hanner:1982,Hanner:1994,Wooden:1999,Harker:2002,Harker:2004err}.  As discussed in \S~\ref{sec:intro}, Mg-rich crystalline silicates are the dominant crystalline silicate component seen in comet C/1995~O1~(Hale-Bopp) \citep{Crovisier:1997,Harker:2002,Lisse:2006Sci} and in comets in general \citep{Hanner:2010}.  Resonant peak locations have been shown by observations \citep{Molster:2002iii} and by laboratory experiments to vary depending on the shape and size of the sampled grains \citep{Min:2005HB, Mutschke:2009, Tamanai:2006ApJ, Tamanai:2009CDNF, Koike:2010}.  As such, we focus our study on crystalline forsterite with polyhedral shapes, and choose optical constants from \citet{Steyer:1974thesis} to model our grains.  The optical constants of \citet{Steyer:1974thesis} were derived from specular reflectance measurements of a polished crystal surface and so inherently do not contain grain shape-dependent effects.  Over the 8 --40~\micron \ wavelength range, the optical constants of \citet{Steyer:1974thesis} have been proven to be equivalent to the optical indices of refraction derived by \citet{Fabian:2001} from IR transmission measurements of polycrystalline forsterite powders (smaller than 1~\micron) prepared by ball-grinding and dispersed in KBr.  Note that the 49~\micron \ and 69~\micron \ features were discerned from the measurements of polycrystalline powders \citep{Fabian:2001}, but these features are beyond the wavelength range of this study.

%%%%%%%%%%%%%%%%%%%%%%%%%%%%%%%%%%%%%%%%%%%%%%%%%%
\section{RESULTS}
\label{sec:results}

%%%%%%%%%%%%%%%%%%%%%%%%%%%%%%%%%%%%%%%%%%%%%%%%%%
\subsection{Crystal Shape} 
\label{sec:results_shape}

We investigate the effects of forsterite crystal shape on both the resonant absorption features' shape and the resonant features' peak location.  The primary features of interest are the strong features located near 10, 10.5, 11, 16, 19, 23, 27, and 33.5~\micron \ and the weaker, yet significant features near 10.5, 11.9, and 25.2~\micron.  The investigation focuses upon understanding how the crystallographic anisotropic nature of forsterite affects the 8~--~40~\micron \ crystalline resonance features with respect to crystallographic axis length and specific grain shape characteristics, such as the importance of the presence of and angles between the vertices, edges, and faces of polyhedral crystal shapes.  We begin by characterizing the spectral trends associated with elongating or reducing one of the crystallographic a-, b-, or c-axes for rectangular prisms.  The rectangular prisms (henceforth referred to as bricks) are the simplest polyhedral shape to examine the effects of elongating or reducing a crystallographic axis for the DDA since they provide a uniform number of dipoles along all the crystallographic dimensions.  The spectral effects of elongation and reduction of bricks thus provide a fiducial for how the balance between crystal shape and length of the crystallographic axes affect the forsterite absorption resonance features.  

In \S~\ref{sec:results_axes_sym}, the absorption feature effects of elongations or reductions of a single crystallographic axis are characterized using the base shape of a cube.  The equal lengths, or symmetry, of all the dimensions of a cube limit the effects on the absorption features due to the other two symmetric axes.  This allows for identification of spectral trends associated with elongating and reducing the a-, b-, and c- crystallographic axes.  However, the equal length symmetry of the two unaltered axes may be its own source of spectral effects.  To test whether this is the case and confirm the spectral trends, we repeat the elongation and reduction experiment of bricks in \S~\ref{sec:results_axes_asym} using a base shape of a triaxial brick, where the two unaltered axes are set to the ratio of 0.8 and 1.2 times the length of the axis to be altered.  We find that asymmetry does induce significant spectral effects, such that the two asymmetries (0.8,1.2 and 1.2,0.8) have divergent spectral features.  Regardless, the spectral trends demonstrated by the symmetric case still hold for each asymmetric case.

The divergent spectral behavior demonstrated by the asymmetric bricks indicates that the specific crystallographic axial ratio is important to the overall absorption characteristics of a forsterite grain.  In  \S~\ref{sec:results_faces_edges}, non-brick shapes with similar axial ratios are compared to demonstrate that the effects of elongations and reductions hold for more complex grain shapes and to provide a method to test the effects of differences in grain shape.  Specifically, the comparison allows for a direct way to investigate the effects of specific grain shape with respect to the presence and angles between faces, edges, and vertices through examination of the absorption feature differences between shapes with exact or very similar axial ratios.  

%%%%%%%%%%%%%%%%%%%%%%%%%%%%%%%%%%%%%%%%%%%%
\subsubsection{Variation of Crystallographic Axes Lengths: The Symmetric Case}
\label{sec:results_axes_sym}

Through understanding the effect of the simplest shape variations, it is possible to understand more complex shape effects.  Hence, we start with elongations and reductions of the symmetric case of the rectangular prism: the cube.  Shapes delineated by faces parallel to crystallographic axes are the simplest to relate to elongations or reductions of the axes lengths of a cube.  Shapes delineated by two (triangular prism) or more faces being non-parallel to the crystallographic axes show similar trends as the elongated cube in the wavelengths and strengths of the features but the feature shapes are not predictable as a linear combination of elongations of a cube.  

The symmetry of a cube provides for the most controlled examination of the spectral effects related to an elongated or reduced single crystallographic axis.  The elongations and reductions are defined by systematically varying the length of one of the crystallographic axes (the a-, b-, or c-axis) with respect to the other two while keeping the size of the grain constant at $a_{\rm eff}$ = 0.1~\micron.  The elongations used are 1.2, 1.4, 1.6, 1.8, 2.0, 2.6, 3.0, and 4.0 times the original length of the cube-shaped crystal.  For example, a 4.0 times crystallographic a-axis grain is essentially four cubes stacked upon each other in the a-axis direction giving an a~:~b~:~c = 4~:~1~:~1 axial ratio.  The reductions used are 0.8, 0.6, 0.4, 0.2, and 0.1 times the original length of a the cube-shaped crystal, such that at large reductions -- 0.2 and 0.1 times -- the grain shape is best described as a platelet.

\begin{figure} ÊÊÊÊÊÊÊÊÊÊ%Fig. 2
ÊÊÊ \begin{center}
ÊÊÊ ÊÊÊ \epsfig{file=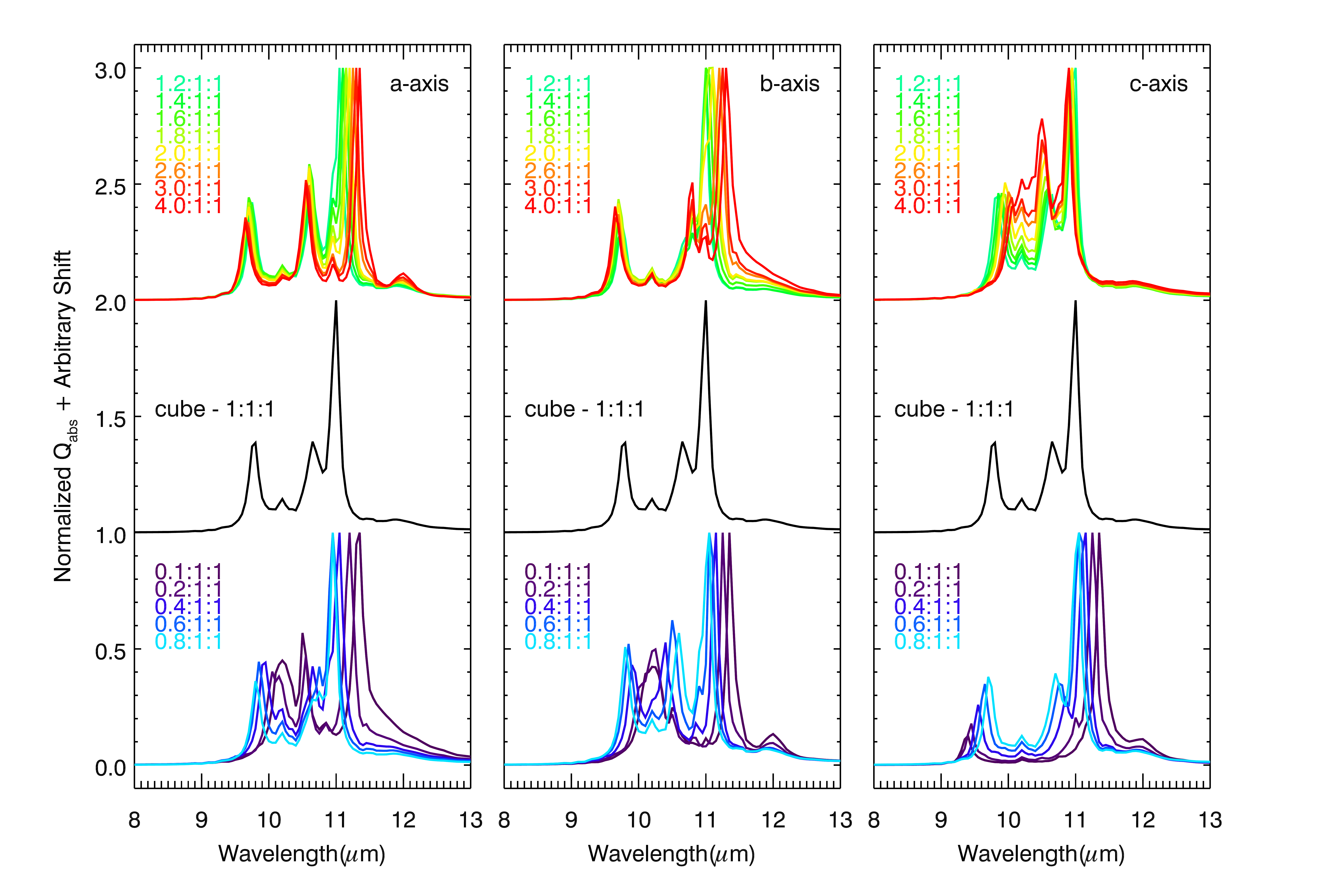,height=5.0in,width=6.0in}
        \caption{The 8~--~13~\micron \ absorption efficiencies normalized to the 11~\micron \ peak for a- (\emph{left panel}), b- (\emph{middle panel}), and c-axis (\emph{right panel}) elongations and reduction of the crystallographic axes with respect to a fiducial cube shape (\emph{black}, middle). The top curves in warm colors are 1.2, 1.4, 1.6, 1.8, 2.0, 2.6, 3.0, and 4.0 times the original length of the cube elongations.  The bottom curves in cool colors are 0.8, 0.6, 0.4, 0.2, and 0.1 times the original length of the cube reductions.}
     \label{fig:axial_ratios_8_13}
ÊÊÊ \end{center}
\end{figure}

\begin{figure} ÊÊÊÊÊÊÊÊÊÊ%Fig. 3
ÊÊÊ \begin{center}
ÊÊÊ ÊÊÊ \epsfig{file=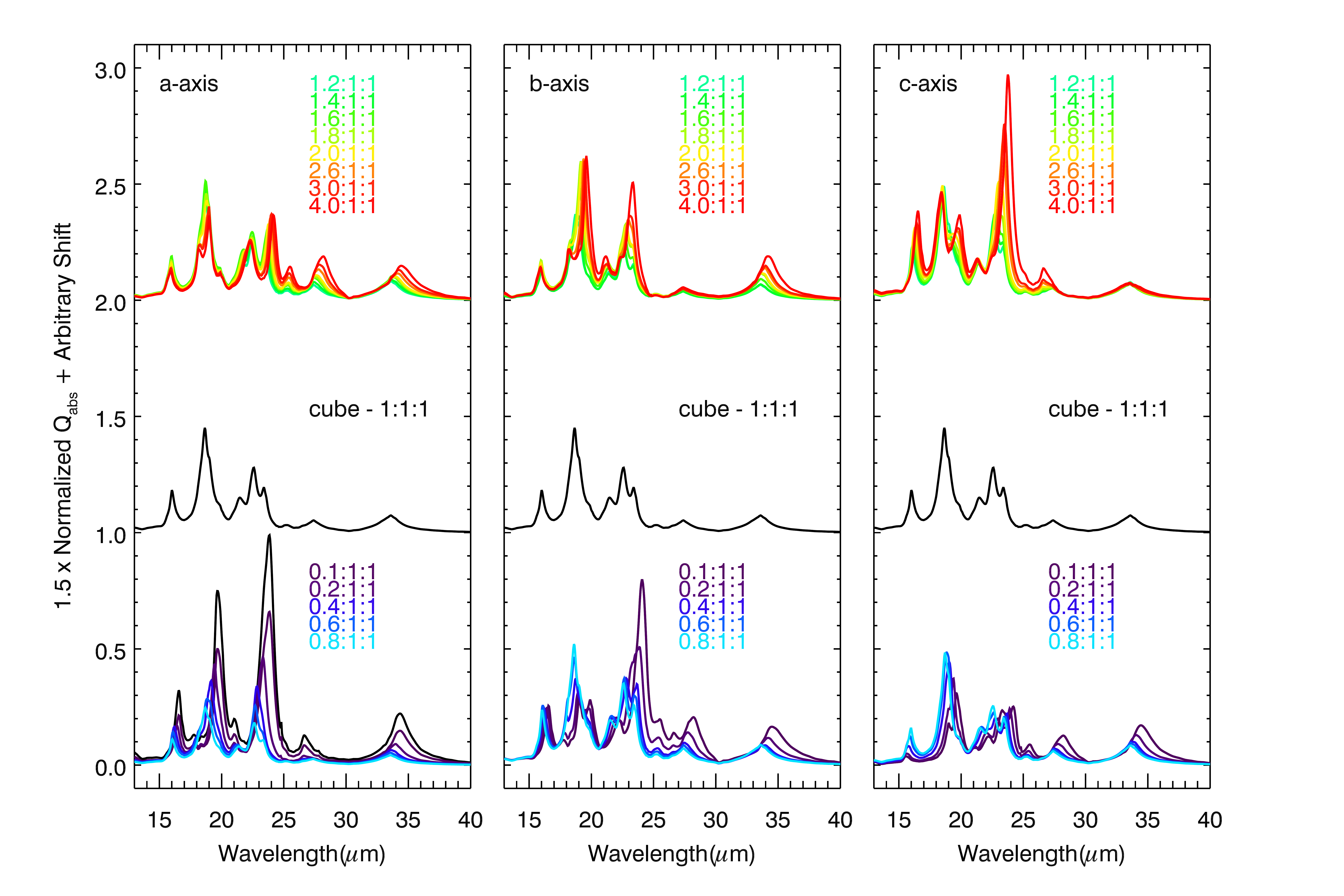,height=5.0in,width=6.0in}
        \caption{The 13~--~40~\micron \ absorption efficiencies normalized to the 11~\micron \ peak for a- (\emph{left panel}), b- (\emph{middle panel}), and c-axis (\emph{right panel}) elongations and reduction of the crystallographic axes with respect to a fiducial cube shape (\emph{black}, middle). The top curves in warm colors are 1.2, 1.4, 1.6, 1.8, 2.0, 2.6, 3.0, and 4.0 times the original length of the cube elongations.  The bottom curves in cool colors are 0.8, 0.6, 0.4, 0.2, and 0.1 times the original length of the cube reductions.  The $Q_{\rm abs}$ values are multiplied by a factor of 1.5 for visual clarity.}
     \label{fig:axial_ratios_13_40}
ÊÊÊ \end{center}
\end{figure}

The absorption efficiencies of the elongations and reductions are displayed in Figs.~\ref{fig:axial_ratios_8_13} and \ref{fig:axial_ratios_13_40} for the 8~--~13~\micron \ and 13~--~40~\micron \ range, respectively.  The fiducial spectrum of the cube is described as follows: the `10~\micron \ feature' and `11~\micron \ feature' appear at 9.8 and 11.0~\micron, respectively, with minor peaks at 10.2 and 10.6~\micron .  The stronger 18 and 23~\micron \ features are asymmetric, such that the 18~\micron \ feature has two shoulders on the long wavelength side of the peak, and the 23~\micron \ is a triple-peaked, `trident'-shaped feature.  The weaker 16, 27, and 33.5~\micron \ features are singular symmetric peaks.

The elongations/reductions of any one axis exhibit consistent spectral trends, i.e., the 2.0 times elongations are intermediary between the 1.0 and 4.0 times  elongations.  The effects of elongations and reductions approach having distinct spectral features as the elongations and reductions become `extreme', i.e., $\ge$~2.6 times elongated or $\le$ 0.2 times reduced.  In these `extreme' cases, we refer to elongated shapes as columns (occasionally referred to as `whiskers' in the literature) and reduced shapes as platelets.

Below, we discuss the specific spectral effects with respect to feature shape, strength, and location for elongations/reductions for each of the a-, b-, and c-axes.

\emph{Elongating the cube along the a-axis} shifts the 10~\micron \ peak to shorter wavelengths and  the 11~\micron \ peak to longer wavelengths (Fig.~\ref{fig:axial_ratios_8_13}, \emph{left panel, top}).  The 10~\micron \ feature also slightly decreases in relative strength with respect to the 11~\micron \ feature (hereafter, all references to changes in relative strength are relative to the 11~\micron \ feature).  For the weaker features in the 8~--~13~\micron \ range, elongating the a-axis causes the 10.6~\micron \ feature to become a distinct separate peak near 10.5~\micron \ that shifts to shorter wavelengths and decreases in feature strength; the 10.2 and 11.9~\micron \ peaks become more distinct with increased relative strengths.  The cube has a 23~\micron \ `trident'-shaped feature that with a-axis elongation morphs into a bifurcated pair of peaks with a weak third peak on the long wavelength end of the feature (Fig.~\ref{fig:axial_ratios_13_40}, \emph{left panel, top}).  This bifurcated pair of peaks is unique to a-axis elongations, and should be discernible in spectroscopic observations. 

\emph{Elongating the cube along the b-axis} shifts the 10~\micron \ peak to shorter wavelengths and the 11~\micron \ peak to longer wavelengths (Fig.~\ref{fig:axial_ratios_8_13}, \emph{middle panel, top}).  For the weaker features in the 8~--13~\micron \ range, elongating the b-axis causes the 
10.6~\micron \ feature to split into two narrow features at 10.8~\micron \ and at 11.0~\micron, with the 10.8~\micron \ increasing and the 11.0~\micron \ decreasing in relative strength.  With increasing b-axis elongation, the 11.9~\micron \ feature becomes an increased absorptivity tail for the red side of the 11~\micron \ feature.  The cube has an 18~\micron \ feature with two long wavelength shoulders that with b-axis elongation fade and a short wavelength-side shoulder develops, becoming a minor companion peak for `extreme' elongations (Fig.~\ref{fig:axial_ratios_13_40}, \emph{middle panel, top}).  The cube has a 23~\micron \ `trident'-shaped feature that with b-axis elongation becomes a well-defined single peak plus a minor peak between the 18 and 23~\micron \ features.  The ability to discern b-axis elongation from a- or c-axis elongations is possible through the absence of the distinct spectral features associated with the a- or c-axis elongations.

\emph{Elongating the cube along the c-axis} shifts the 10~\micron \ feature to longer wavelengths for non-`extreme' elongations.  For `extreme' elongations, the 10~\micron \ feature becomes a shelf feature that intersects with the 10.6~\micron \ feature.  As elongation increases, the 10.6~\micron feature shifts to 10.5~\micron \ and becomes a distinct peak that is nearly equal in strength to the 11~\micron \ feature (Fig.~\ref{fig:axial_ratios_8_13}, \emph{right panel, top}).  The 11~\micron \ feature shifts to slightly shorter wavelengths, which is the opposite behavior of the a- and b-axis elongations.  With c-axis elongation, the 16~\micron \ feature significantly increases in relative strength and the 18~\micron \ feature morphs from a single peak into a bifurcated, double peak.  The cube has a 23~\micron \ `trident'-shaped feature that with c-axis elongation becomes a single prominant peak with increasing strength and shifts slightly toward longer wavelengths; by 4.0 times elongation the 23~\micron \ feature is nearly as strong (approximately two-thirds) as the 11~\micron feature (Fig.~\ref{fig:axial_ratios_13_40}, \emph{right panel, top}).  With c-axis elongation, the 33.5~\micron \ feature does not change in feature shape, location, or strength.  The c-axis elongation behavior for the 16 and 33.5~\micron \ features is the opposite for what is observed for a- and b-axis elongations.  The 27~\micron \ feature also increases in relative strength and develops a distinct `shark-tooth' shape, rising quickly and falling slowly.  The c-axis `extreme' elongations produce the combination 10~\micron \ shelf -- 10.6~\micron~-- 11~\micron \ feature, significantly increased 23~micron\ feature strength, and a `shark-tooth' 27~\micron \ feature, which are distinct spectral patterns for c-axis elongations compared to a- and b-axis elongations that should be discernible in observations. 

In contrast to elongations, where one axis becomes increasingly longer over the other two, the case of reductions is where one axis becomes shorter than the other two.  As exhibited by the spectra at the bottom of Figs.~\ref{fig:axial_ratios_8_13} and \ref{fig:axial_ratios_13_40}, axial reductions with one axis significantly shorter than the other two do not follow the same trends established with elongations.  In general, reductions create more complex spectral feature trends with increasing reduction, which can be considered a combination of elongations of the two other (unaltered) axes.  

\emph{Reducing the cube along the a-axis} from 0.8 times to 0.1 times the original length of the cube shifts the 10 and 11~\micron \ features to longer wavelengths (Figs.~\ref{fig:axial_ratios_8_13} \emph{left panel, bottom}).  The 10.2~\micron \ feature increases in relative strength and merges with the 10~\micron \ feature for `extreme' reductions ($\le 0.2$ times) creating a rounded peak.  Also at `extreme' reductions, the cube's 10.6~\micron \ feature moves to shorter wavelengths and becomes a narrow peak at 10.5~\micron \ with increasing feature strength.  The combination of the 10.2 and 10.6~\micron \ features' shape and location is similar to that of c-axis elongation.  However, at `extreme' reductions, the 11~\micron \ feature develops an absorption tail that is similar to b-axis elongations.  The 16, 18, 23, and 33.5~\micron \ features all shift to significantly longer wavelengths ($\Delta\lambda \lesssim$~1.0~\micron) and become relatively stronger with increasing a-axis reduction; these feature shifts are characteristic of b- and c-axis elongations.  A-axis reductions also create a significant increase in the relative strength of the 23~\micron \ feature, and a `shark-tooth'-shaped 27~\micron \ feature, which are characteristic of c-axis elongations.  The overall feature structure suggests that the spectral features for a-axis reductions are in reality a combination of b- and c-axis elongation spectral features.  The pattern of axial reductions of one crystallographic axis having spectral characteristics similar to a combination of elongations of the two unaltered axes holds for a-, b- and c-axis reductions.

\emph{Reducing the cube along the b-axis} shifts the 10 and 11~\micron \ features to longer wavelengths.  Similar to a-axis elongations, b-axis reductions have a distinct 10.6~\micron \ feature for all non-`extreme' reductions ($\ge 0.4$ times).  Towards `extreme' b-axis reductions, the 10 and 10.6~\micron \ features become a single, rounded peak centered near 10.2~\micron.  With b-axis reductions, the minor 11.9~\micron \ feature becomes increasingly strong, which is a characteristic of a-axis elongations.  With b-axis reductions, the 16, 18, 23, 27, and 33.5~\micron \ features all shift to longer wavelengths.  Also, the 18 and 23~\micron \ features significantly change in feature shape.
The cube has a 18~\micron \ single peak with two long wavelength shoulders that with b-axis reductions morphs into a bifurcated double-peak that decreases in relative strength.   The cube has a 23~\micron \ `trident'-shaped feature that morphs to a singular peak with significantly increasing relative strength. The 18~\micron \ bifurcated double-peak and the strong 23~\micron \ singular peak are both characteristics of c-axis elongations. 

\emph{Reducing the cube along the c-axis} shifts the 10~\micron \ feature to shorter wavelengths and shifts the 11~\micron \ feature to longer wavelengths.  C-axis reductions are the only axial reductions that shift the 10~\micron \ feature to shorter wavelengths, which is opposite the characteristic of both a- and b-axis elongations where this feature shifts to shorter wavelengths.  As the c-axis is reduced, the 10~\micron \ feature also significantly decreases in relative strength, and the 10.6~\micron \ feature diminishes to a small peak at 11.0~\micron.  Neither of these spectral changes are apparent in a- or b-axis elongations suggesting that not all spectral characteristics associated with a single axis reduction can be understood as a combination of elongations of the two unaltered axes.  With c-axis reductions, the 18, 23, 27, and 33.5~\micron \ features shift to longer wavelengths; the 16~\micron \ feature shifts to shorter wavelengths and significantly diminishes in relative strength.  
The cube has an 18~\micron \ feature with two long wavelength shoulders that with c-axis reductions morphs into a singular peak with a minor sub-peak on the short wavelength-side.    
The cube has a 23~\micron \ `trident'-shaped feature that morphs into to a stair-stepped-shaped triple peak of increasing strength at longer wavelengths.  The change in the 18~\micron \ feature's shape is similar to b-axis elongations, but the change in the 23~\micron \ feature's shape is not apparent in either a- or b-axis elongations.

The more complex behavior for crystallographic axis reductions largely can be understood as a combination of elongations of the two unaltered axes.  As the target axis is reduced, the other two axes becoming significantly longer, or elongated, in comparison.  These combinations create distinct spectral features for a-, b-, and c-axis elongations that should be observationally identifiable in the spectra of astrophysical objects.

In this section, the spectral changes presented for the symmetric case of the cube demonstrate that elongations and reductions of each crystallographic axis have distinct spectral characteristics over the 8~--~40~\micron \ spectral range.  The trends with increasing elongation and reduction are predictable and different for each crystallographic axis.  To summarize: 
The \emph{a-axis elongations} have the distinct 10.5~\micron \ feature and a bifurcated 23~\micron \ double-peak.
The \emph{a-axis reductions} trend toward a 10.2~\micron \ rounded peak paired with a narrow 10.5~\micron \ peak, and with increasing a-axis reduction the 15~--~40~\micron \ features all move to significantly longer wavelengths and the 23~\micron \ feature strength becomes significantly enhanced.
The \emph{b-axis elongations} have spectral features that are singular peaks, exhibit a 10.8~\micron \ shelf, and an absorptive tail on the long wavelength side of the 11~\micron \ feature.
The \emph{b-axis reductions} trend toward a singular, broad 10.2~\micron \ feature, and an enhancement in the 23~\micron \ feature strength that is slightly weaker than that of a-axis reductions.  
The \emph{c-axis elongations} trend toward a 10~\micron \ shelf, a bifurcated 18~\micron \ feature, and a 23~\micron \ feature that is significantly enhanced in relative strength.
The \emph{c-axis reductions} exhibit a 10~\micron \ feature that diminishes in relative strength and a `stair-stepped' 23~\micron \ feature.
In the case of `extreme' elongations ($\ge$~2.6 times the original length of the cube, also called `columns'), and reductions ($\le$ 0.2 times the original length of the cube, also called `platelets'), the spectral features are distinct enough such that they should be discernible in observations.

%%%%%%%%%%%%%%%%%%%%%%%%%%%%%%%%%%%%%%%%%%%%
\subsubsection{Variation of Crystallographic Axes Lengths: The Asymmetric Case}
\label{sec:results_axes_asym}

The previous section detailed the spectral effects due to the elongation/reduction of a single crystallographic axis with respect to the base shape of a cube.  Here, we repeat the elongation/reduction exercise with an asymmetric rectangular prism (denoted as an asymmetric brick) with each of the three edges having differing lengths.  

The shapes for this exercise of the elongations/reductions of the asymmetric case have edge lengths of 0.8$l$, 1.0$lX$, and 1.2$l$, where $l$ is the edge length of the cube (in dipoles) in the previous section, and $X$ is the multiplicative factor for the elongations and reductions. Since the edge length of the cube, $c$, is in every term, we drop it from the notation.  For the `fiducial bricks' ($X$ = 1), there are six asymmetric realizations that include: two a-axis bricks, one with a longer b-axis (denoted a~:~b~:~c = 1.0~:~1.2~:~0.8) and one with a longer c-axis (a~:~b~:~c = 1.0~:~0.8~:~1.2); two b-axis bricks (0.8~:~1.0~:~1.2 and 1.2~:~1.0~:~0.8); and two c-axis bricks (0.8~:~1.2~:~1.0 and 1.2~:~0.8~:~1.0).  Similarly, for 2 times elongations, the six asymmetric renditions include shapes for the a-axis (2.0~:~0.8~:~1.2 and 2.0~:~1.2~:~0.8), b-axis (0.8~:~2.0~:~1.2 and 1.2~:~2.0~:~0.8), and c-axis (0.8~:~1.2~:~2.0 and 1.2~:~0.8~:~2.0).  

\begin{figure} ÊÊÊÊÊÊÊÊ%Fig. 4
ÊÊÊ \begin{center}
ÊÊÊ ÊÊÊ \epsfig{file=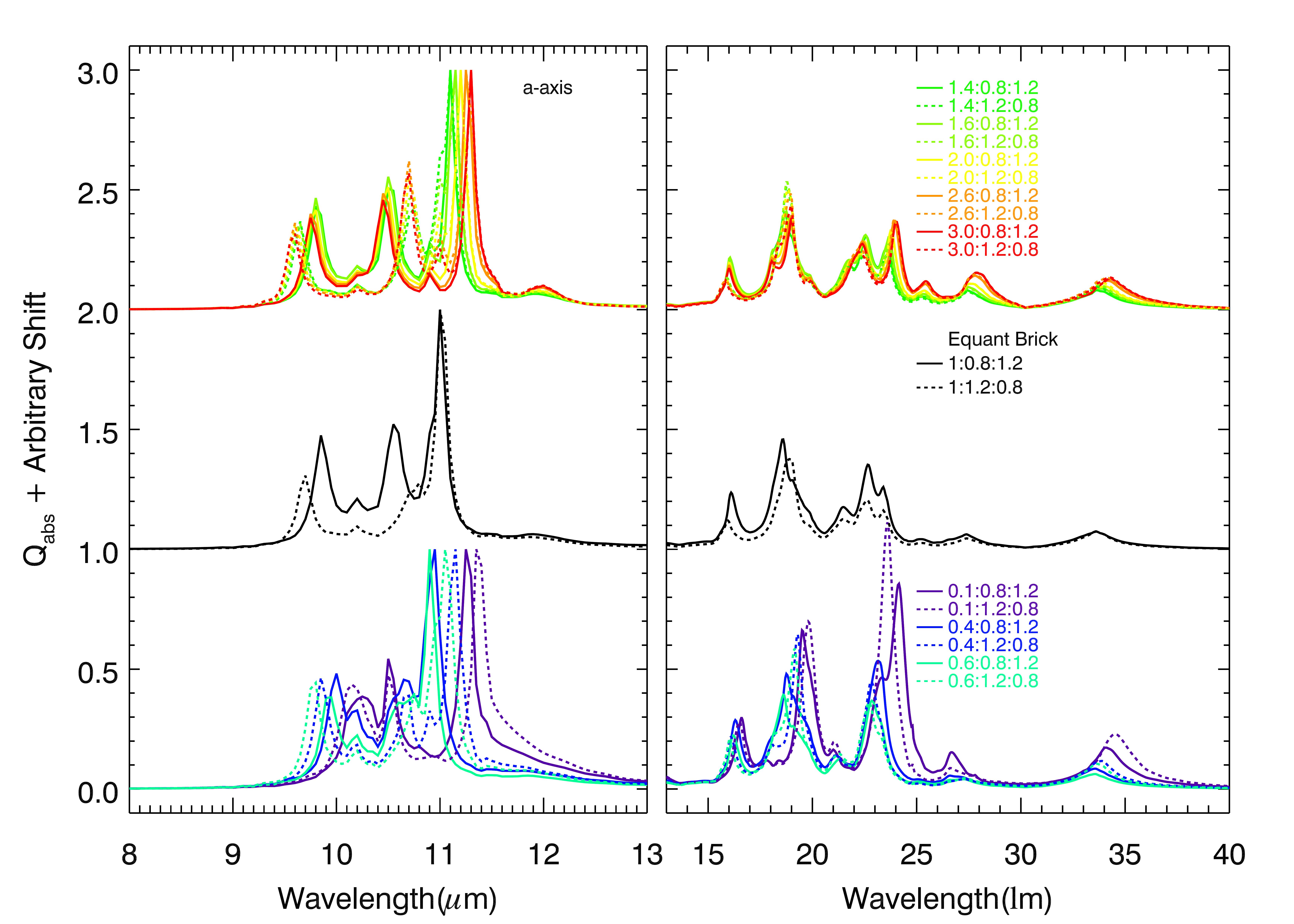,height=5.0in,width=6.0in}
        \caption{The 8~--~13~\micron \ (\emph{left panel}) and 13~--~40~\micron \ (\emph{right panel}) absorption efficiencies normalized to the 11~\micron \ peak for the a-axis elongations (\emph{top curves}) and reductions (\emph{bottom curves}) for tri-axially asymmetric bricks compared to `fiducial' bricks with axial ratios 1.0~:~1.2~:~0.8 and 1.0~:~0.8~:~1.2 (\emph{middle curves}). The \emph{dashed} curves have the b-axis as the second longest (1.2) axis, and the \emph{solid} curves have the c-axis as the second longest axis.  The elongations are 1.6, 2.0, 2.6, and 3.0 times the 1.0 side of the `fiducial' bricks, and the reductions are 0.6, 0.4, and 0.1 times the 1.0 side of the `fiducial' bricks.  The 13~--40~\micron \ $Q_{\rm abs}$ values are multiplied by a factor of 1.5 for visual clarity.}
     \label{fig:asymmetric_bricks_a}
ÊÊÊ \end{center}
\end{figure}

\begin{figure} ÊÊÊÊÊÊÊÊ%Fig. 5
ÊÊÊ \begin{center}
ÊÊÊ ÊÊÊ \epsfig{file=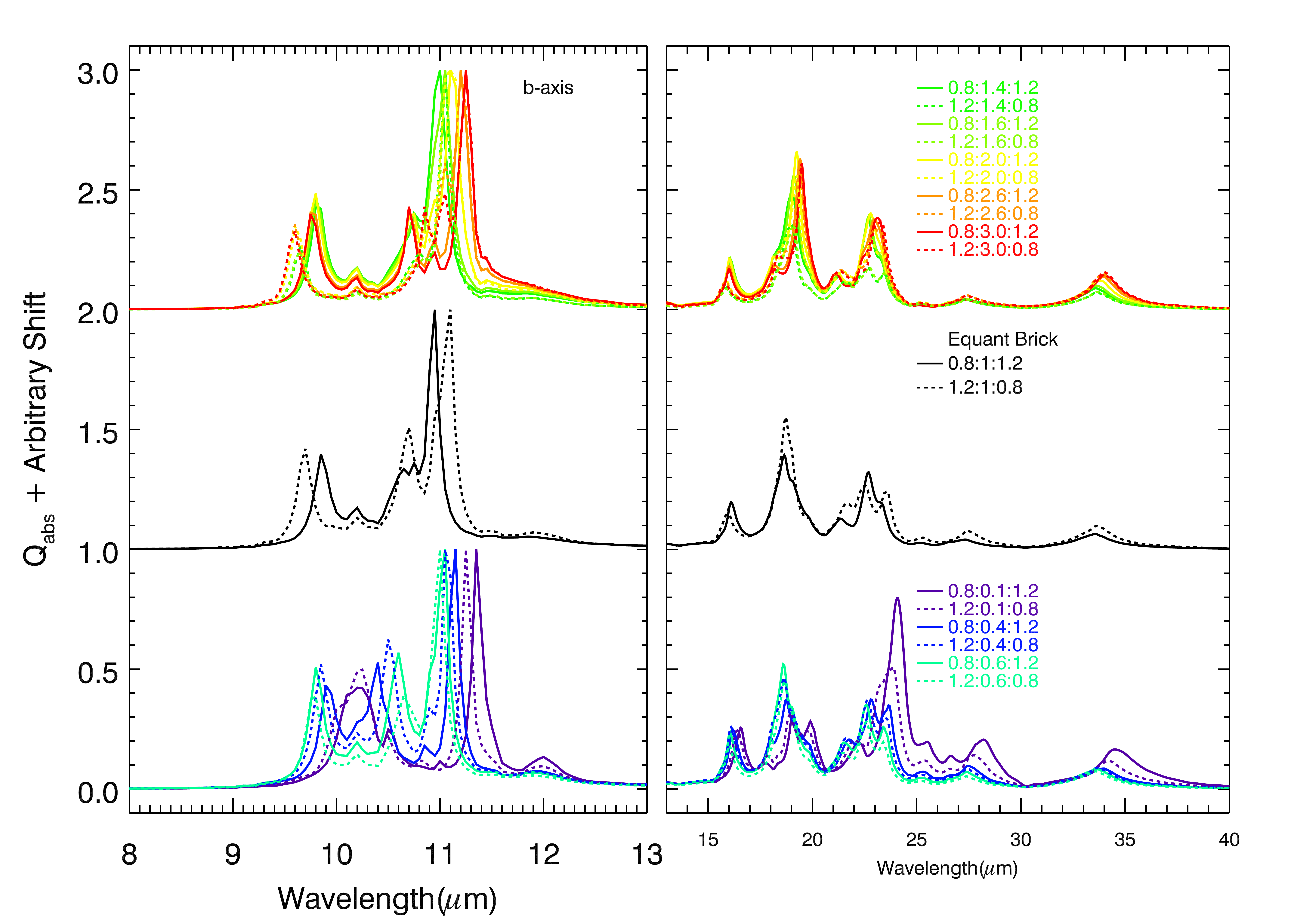,height=5.0in,width=6.0in}
        \caption{The 8~--~13~\micron \ (\emph{left panel}) and 13~--~40~\micron \ (\emph{right panel}) absorption efficiencies normalized to the 11~\micron \ peak for the b-axis elongations (\emph{top curves}) and reductions (\emph{bottom curves}) for tri-axially asymmetric bricks compared to `fiducial' bricks with axial ratios 0.8~:~1.0~:~1.2 and 1.2~:~1.0~:~0.8 (\emph{middle curves}). The \emph{dashed} curves have the a-axis as the second longest (1.2) axis, and the \emph{solid} curves have the c-axis as the second longest (1.2) axis.  The elongations are 1.6, 2.0, 2.6, and 3.0 times the 1.0 side of the `fiducial' bricks, and the reductions are 0.6, 0.4, and 0.1 times the 1.0 side of the `fiducial' bricks.  The 13~--40~\micron \ $Q_{\rm abs}$ values are multiplied by a factor of 1.5 for visual clarity.}
     \label{fig:asymmetric_bricks_b}
ÊÊÊ \end{center}
\end{figure}

\begin{figure} ÊÊÊÊÊÊÊÊ%Fig. 6
ÊÊÊ \begin{center}
ÊÊÊ ÊÊÊ \epsfig{file=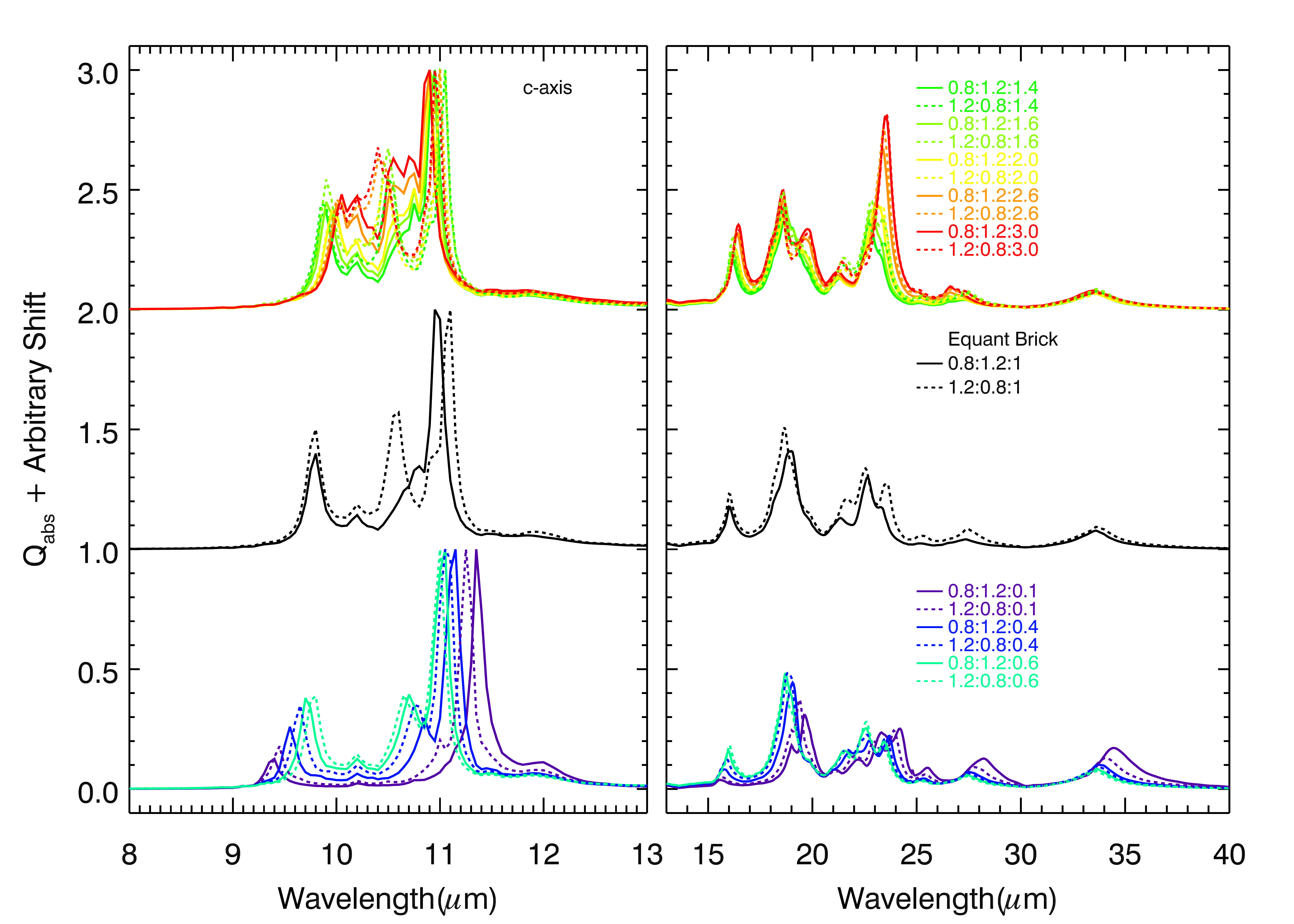,height=5.0in,width=6.0in}
        \caption{The 8~--~13~\micron \ (\emph{left panel}) and 13~--~40~\micron \ (\emph{right panel}) absorption efficiencies normalized to the 11~\micron \ peak for the c-axis elongations (\emph{top curves}) and reductions (\emph{bottom curves}) for tri-axially asymmetric bricks compared to `fiducial' bricks with axial ratios 0.8~:~1.2~:~1.0 and 1.2~:~0.8~:~1.0 (\emph{middle curves}). The \emph{dashed} curves have the a-axis as the second longest (1.2) axis, and the \emph{solid} curves have the b-axis as the second longest (1.2) axis.  The elongations are 1.6, 2.0, 2.6, and 3.0 times the 1.0 side of the `fiducial' bricks, and the reductions are 0.6, 0.4, and 0.1 times the 1.0 side of the `fiducial' bricks. The 13~--40~\micron \ $Q_{\rm abs}$ values are multiplied by a factor of 1.5 for visual clarity.}
     \label{fig:asymmetric_bricks_c}
ÊÊÊ \end{center}
\end{figure}

Figures~\ref{fig:asymmetric_bricks_a}, \ref{fig:asymmetric_bricks_b}, and \ref{fig:asymmetric_bricks_c} show the a-, b-, and c-axis elongations/reductions, respectively.  Each figure shows $X = 1.2-3.0$ elongations (\emph{top spectra}), the fiducial asymmetric bricks (\emph{middle spectra}), and $X = 0.8-0.1$ reductions (\emph{bottom spectra}). The fiducial asymmetric bricks (\emph{middle}) are analogous to the fiducial cube in Figs.~\ref{fig:axial_ratios_8_13} and \ref{fig:axial_ratios_13_40} (\emph{middle}).  
For each elongation/reduction of a specific crystallographic axis, \emph{solid lines} and \emph{dashed lines} represent the two different asymmetric renditions (0.8~:~1.2 and 1.2~:~0.8, pertaining to the two unaltered axes).  
Certain spectral features exhibit distinguishable differences between the two asymmetric renditions and we denote these features as being `sensitive to the effects of asymmetry'.  Other spectral features are indistinguishable for the two asymmetric renditions and are deemed `insensitive to the effects of asymmetry'.  Comparing and contrasting the asymmetric brick elongations/reductions in Figs.~\ref{fig:asymmetric_bricks_a}, \ref{fig:asymmetric_bricks_b}, and \ref{fig:asymmetric_bricks_c}  with the symmetric cube elongations/reductions in Figs.~\ref{fig:axial_ratios_8_13} and \ref{fig:axial_ratios_13_40}, shows: 1)~the trends in feature shape, peak location and relative strength that exist for the elongations/reductions of the cube also hold for elongations/reductions of \emph{both} asymmetric renditions of the brick; and 2)~that certain, but not all, spectral features of a given elongation/reduction of a specific axis exhibit differences in their feature shape, peak location, and relative strength between the two asymmetric renditions (0.8~:~1.2 \emph{versus} 1.2~:~0.8 for the axes that are not elongated/reduced).

For the \emph{a-axis elongations of the asymmetric brick} (Fig.~\ref{fig:asymmetric_bricks_a}, \emph{top}), the 10.6~\micron \ feature in the symmetric case (of the cube, Fig.~\ref{fig:axial_ratios_8_13}) `splits' (bifurcates) into a 10.4~\micron \ feature for the longer c-axis asymmetric rendition (0.8b~:~1.2c) and a 10.6~\micron \ feature for longer b-axis asymmetric rendition (1.2b~:~0.8c); the 11~\micron \ feature is insensitive to asymmetry, i.e., the \emph{dashed} and \emph{solid} lines plot over one another.  Furthermore, with a-axis elongations, the 11~\micron \ feature shifts to longer wavelengths and the 10.4 or 10.6~\micron \ features only slightly shift to shorter wavelengths and slightly decrease in relative strength, mirroring the a-axis cube-elongations (Fig.~\ref{fig:axial_ratios_8_13}, \emph{top}). 

For the \emph{b-axis elongations of the asymmetric brick} (Fig.~\ref{fig:asymmetric_bricks_b}, \emph {top}), the 9.7~\micron \ feature in the symmetric case splits into 9.6~\micron \ for longer a-axis asymmetric rendition and 9.8~\micron \ for the longer c-axis asymmetric rendition; the 11~\micron \ feature is insensitive to asymmetry.  

For the \emph{c-axis elongations of the asymmetric brick} (Fig.~\ref{fig:asymmetric_bricks_c}, \emph{top}), the features shortward of 11~\micron \ show more complicated behavior than the a- or b-axis elongations and are divergent for the two asymmetric renditions (0.8a~:~1.2b and 1.2a~:~0.8b): 
for `extreme' elongations, longer a-axis renditions (1.2a~:~0.8b~:~$X$c, \emph{dotted lines}) develop 10.5~\micron \ peaks whereas longer b-axis renditions (0.8a~:~1.2b~:~$X$c, \emph{solid lines}) maintain a trough.  For c-axis elongations, the 11~\micron \ feature is sensitive to asymmetry.  Despite the variation in sensitivity to asymmetric renditions, all of the trends in spectral feature peak location shifts are equivalent to the symmetric case and the distinguishing features associated with elongations/reductions are present (\S~\ref{sec:results_axes_sym}, last paragraph).  

The spectral characteristics of the elongations with respect to the two asymmetric renditions in Figs.~\ref{fig:asymmetric_bricks_a}, \ref{fig:asymmetric_bricks_b}, and \ref{fig:asymmetric_bricks_c} can be characterized by their primary axis elongation/reduction spectral feature characteristics coupled with spectral characteristics related to {\em elongations of the second longest (1.2) axis}.  As already mentioned, for the $X \ge 2.6$ c-axis elongations in Fig.~\ref{fig:asymmetric_bricks_c}, the a- (\emph{dashed}) and b-axis (\emph{solid}) asymmetric renditions (1.2a~:~0.8b and 0.8a~:~1.2b) exhibit significantly different absorption characteristics between the 10 and 11~\micron \ features.  The longer a-axis rendition of the c-elongated asymmetric brick has a distinct feature at 10.5~\micron \ and a trough between the 10.5~\micron \ feature and the 11~\micron \ feature, which is similar to the a-axis cube-elongation's distinct 10.5~\micron \ peak.  The longer b-axis rendition of the c-elongated asymmetric brick has a broad feature centered near 10.1~\micron \ and a shelf centered near 10.7~\micron \ that connects with the 11~\micron \ feature, which is similar to the b-axis moderate cube-elongation's shelf feature near 10.8~\micron.

Feature sensitivity or insensitivity to asymmetry depends upon the specific axis that is being elongated/reduced.  The variability in sensitivity can be understood through comparing symmetric elongations to the asymmetric elongations, as is demonstrated with the following two examples.  The \emph{11~\micron \ feature} is sensitive to asymmetry for a- and b-axis elongations, but not for c-axis elongation.  Comparison to the elongations in the symmetric case indicates that the c-axis does not play a dominant role in determining the wavelength position of the 11~\micron \ feature, since in the symmetric case c-axis elongation only shifts the 11~\micron \ feature to shorter wavelengths by 0.1~\micron \ over all elongations.  For a- and b-axis elongations, however, the 11~\micron \ feature shifts by $\approx$~0.4~\micron \ in the symmetric case, and hence whether the a- or b-axis is longer in the asymmetric case plays a strong role in determining where the 11~\micron \ feature peak is located.  As with the 11~\micron \  feature, the {\emph 16~\micron \ feature} is sensitive to asymmetry for a- and b-axis elongation (Figs.~\ref{fig:asymmetric_bricks_a} and \ref{fig:asymmetric_bricks_b}, \emph{solid vs.\ dashed} lines), but not for c-axis elongation.  The symmetric elongations show that the relative strength of the 16~\micron \ feature only significantly changes for c-axis elongations.   Hence, the relative strength of the 16~\micron \ feature for a- and b-axis elongations depends on whether the c-axis is the second longest axis.  The difference between this example and the previous one is that here the c-axis is important for the relative strength of the 16~\micron \ feature, while for the previous example the c-axis is relatively unimportant in determining the position of the 11~\micron \ feature.   A general rule is if a certain spectral resonance is sensitive to only a-, b-, or c-axis elongation/reduction and not the other two, then that feature will be sensitive to asymmetry for renditions that include it as a second longest axis.  If the spectral resonance is sensitive to multiple axial elongations/reductions, then the feature will most likely be sensitive to asymmetry in all axial elongation/reduction cases, but to varying degree (e.g, the 10.5, 19, and 23.5~\micron \ features).  

Consider the fact that the spectral differences between the two asymmetric renditions for a specific crystallographic axis elongation/reduction do not systematically increase or decrease with increased elongation/reduction.  For example, for a-axis elongations, the wavelengths of the split features do not change significantly with increased elongation for the (9.6, 9.8~\micron) and (10.7, 10.5~\micron) features for (b-,c-) second longest axis, respectively.  The expectation is that if elongation/reduction effects solely dominate the spectral features' shape, peak location, and size, that with increased elongation or reduction, the effects of asymmetry should lessen.  This, however, is not the case, as the effects of the second longest axis are apparent in the most extreme elongations and reductions.  The failure for the asymmetric effects to dampen out with increased elongations indicates that the elongations/reductions are the primary shape effects on the 8~--~40~\micron \ spectral features and the effects of asymmetry must still be considered.  Hence, the asymmetric triaxial shapes of forsterite crystals are important shape characteristics with respect to spectral feature's shape, peak position, and relative strength.  DDA computations allow explorations of asymmetric shapes or `crystallographically anisotropic shape' \citep{Takigawa:2009}.  Since triaxial asymmetric shapes probably exist in lab samples and astronomical sources, the varying resonant features' sensitivities to the asymmetric renditions of the brick highlight the need to account for the asymmetric triaxial nature of forsterite in modeling and interpreting observational data. 
	
%%%%%%%%%%%%%%%%%%%%%%%%%%%%%%%%%
\subsubsection{Prismatic and Complex Polyhedral Grain Shapes}
\label{sec:results_faces_edges}

Grain shapes up to this point have been limited to bricks (rectangular prisms), which have been employed to identify  elongation/reduction and asymmetry of forsterite as important grain shape characteristics that affect the spectral features' shape, peak location, and relative strength.  Brick shapes have each of their faces parallel to one of the crystallographic axes.  Forsterite crystals in astrophysical environments, however, may not have their faces aligned in this manner.  In a study of the crystalline silicate grains, including forsterite, found in \emph{Stardust} tracks, the grain shapes are reported as euhedral, or rather, polyhedral shapes that have well-defined, sharp crystalline faces \citep{Nakamura:2012}.

 In this section, the DDA modeling exercises are extended to include non-brick polyhedral shapes including other prisms (triangular and hexagonal), pyramids (tetrahedra and rectangular pyramids or octahedra), and dipyramids\footnote{Dipyramidal polyhedra are equivalent to bipyramidal polyhedra.} (elongated triangular dipyramid, elongated hexagonal dipyramid, and dipyramidal brick).  The same axial ratios are set across the grain shapes, which controls the effects of elongation/reduction and asymmetry.  Hence, the observed spectral differences between sets of grain shapes with the same axial ratio are due to differences in the external shape.  

\begin{figure}[!h]ÊÊÊ  %Fig. 7
ÊÊÊ \begin{center}
ÊÊÊ ÊÊÊ \epsfig{file=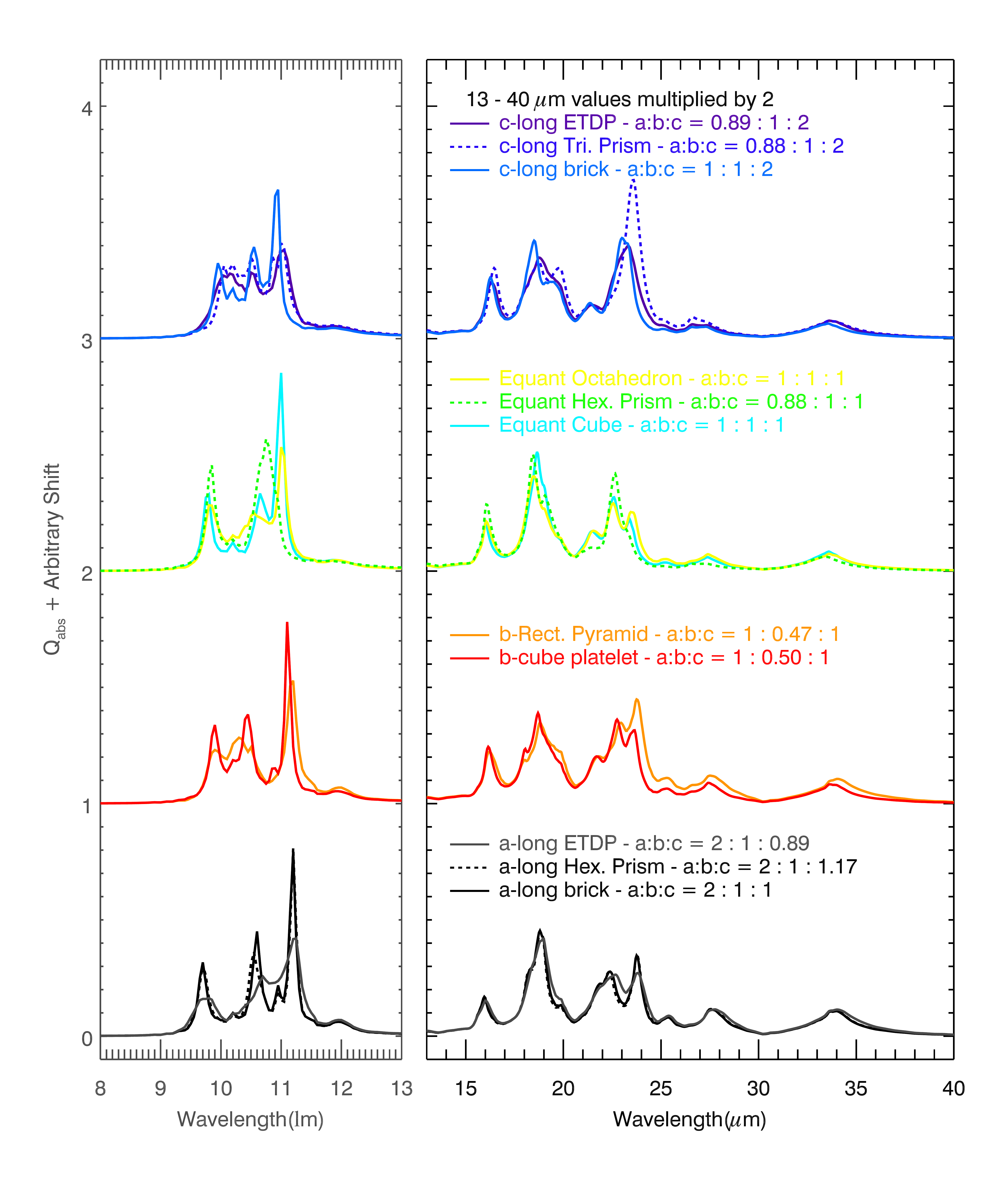,height=5.0in,width=6.0in}
        \caption{The 8~--~40~\micron \ normalized to the 11~\micron \ feature peak absorption efficiencies for rectangular prism (bricks) and non-brick shapes that have similar axial ratio.  The axial ratios from top to bottom are approsimately a~:~b~:~c =  (1~:~1~:~2), (1~:~1~:~1), (1~:~0.5~:~1), (2~:~1~:~1).  Each set of curves contains a brick (\emph{solid curves}) and a (di)pyramidal shape (\emph{dashed curves}).  All sets except the (1~:~0.5~:~1) also contain a non-brick prismatic shape.  The 13~--~40~\micron \ $Q_{\rm abs}$ values are multiplied by a factor of 2 for visual clarity.  The shape abbreviations are ETDP - Elongate Triangular Dipyramd; Tri. - Triangular; DiPyr - Dipyramidal; and Rect. - Rectangular; and Hex. - Hexagonal.}ÊÊÊ ÊÊÊ 
ÊÊÊ \label{fig:brick2nonbrick}
ÊÊÊ \end{center}
\end{figure}

The 8~--~40~\micron \ spectral characteristics for sets of brick and non-brick shapes with similar crystallographic axial ratios are displayed in Fig.~\ref{fig:brick2nonbrick}.  For each set of shapes with similar axial ratios, all of the spectral features show remarkable similarities in feature shape, peak position, and relative strength, and each set is reflective of its elongation/reduction characteristics.  Specifically, the central wavelengths, widths, and general asymmetries of each feature or feature complex are similar.  Spectral differences, however, do exist, and they are primarily for the 10, 11, and 23~\micron \ features; from brick to non-brick the 11~\micron \ feature shifts to longer wavelengths with the exception of the cube to hexagonal prism.  Also, the shapes of the 23~\micron \ features vary.  For the cube to octahedron and from the symmetric b-platelet to b-rectangular pyramid, the feature changes from trident to stair-stepped.  For the c-axis elongated (c-long) brick to the c-axis triangular prism, the 23~\micron \ feature becomes stronger and peaks at a longer wavelength; note that the number of dipoles along the c-axis is greater for the triangular prism than for the elongated triangular prism because of the tips being incorporated into the c-axis elongation.  Generally, the spectral differences between brick and non-brick shapes can be described as the non-brick crystalline resonances having weaker, rounder, and broader 10 and 11~\micron \ features.  These spectral differences suggest that external grain shapes with faces that are not solely parallel to the crystallographic axes are important shape characteristics.  Moreover, these differences are typically greater between prismatic shapes and (di)pyramidal shapes suggesting that the presence of pyramidal structures is potentially an important shape characteristic.

\begin{figure}[!h]	%Fig. 8
ÊÊÊ \begin{center}
ÊÊÊ ÊÊÊ \epsfig{file=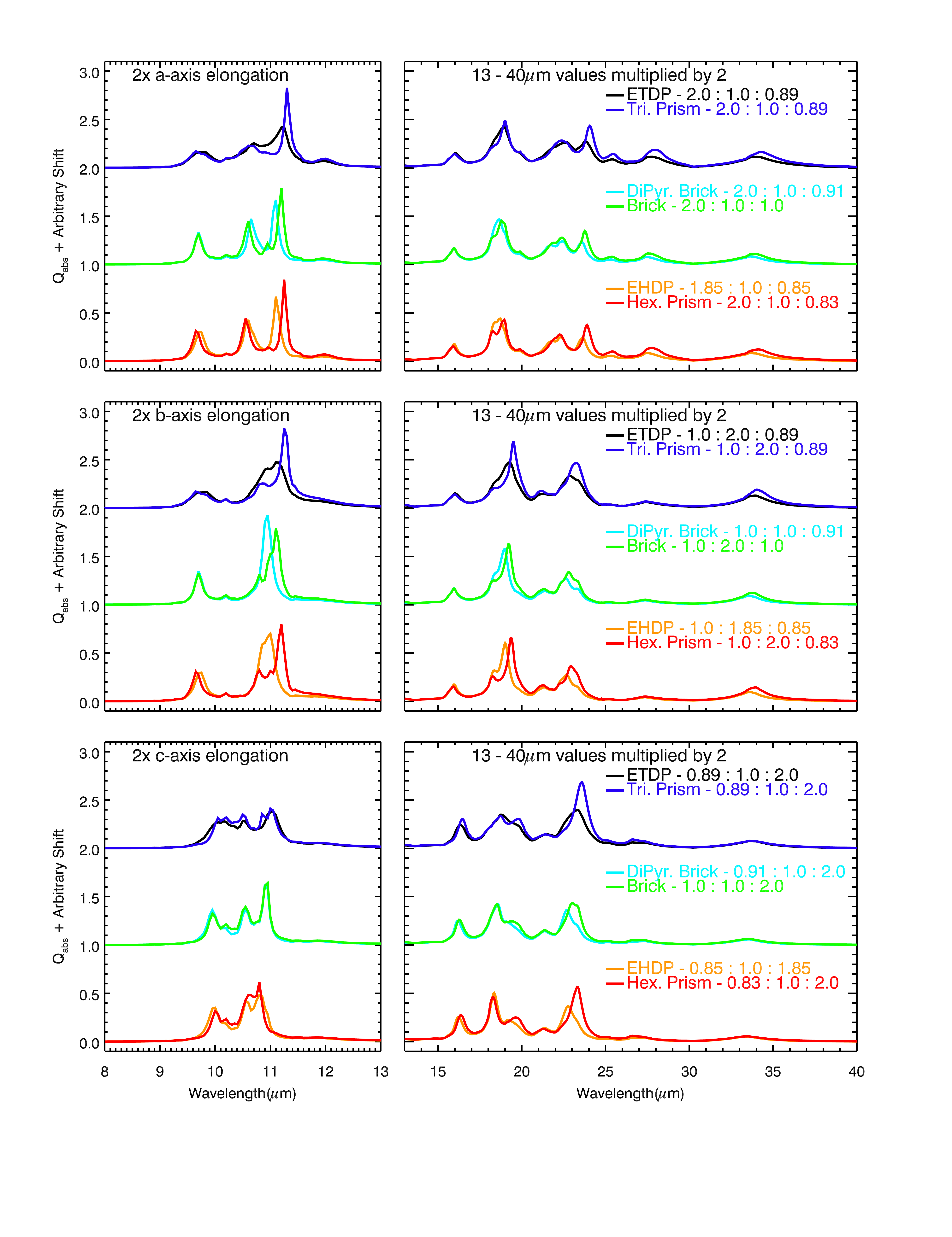,height=6.5in,width=6.5in}
        \caption{The 8~--~40~\micron \ absorption efficiencies for $2~\times$ elongated prismatic shapes and their dipyramidal pair.  The \emph{top panels} are a-axis elongation, the \emph{middle panels} are b-axis elongation, and the \emph{bottom panels} are c-axis elongation.  The prism~--~dipyramid paris are from top-to-bottom in each panel: emph{blues} triangular prism -- elongated triangular dipyramid (ETDP); \emph{blue-greens} brick -- dipryamidal (DiPyr.) brick; and \emph{reds} hexagonal prism -- elongated hexagonal prism (EHDP).  The 13~--40~\micron \ $Q_{\rm abs}$ values are multiplied by a factor of 2 for visual clarity.}ÊÊÊ ÊÊ
ÊÊÊ \label{fig:2times_elongated}
ÊÊÊ \end{center}
\end{figure}

The effects of including pyramidal tips can be constrained through a comparison of the spectral features for prismatic shapes to their dipyramidal pairs of similar axial ratios, e.g., comparing triangular prisms to triangular dipyramids and to elongated triangular dipyramids (triangular dipyramids with trunks) with the same axial ratios.  Such a comparison is shown in Fig.~\ref{fig:2times_elongated} for 2~$\times$~elongated shapes for the a- (\emph{top}), b- (\emph{middle}), and c-axis (\emph{bottom}) for the 8~--~40~\micron \ spectral range.  The prism -- dipyramidal pairs presented in Fig~\ref{fig:2times_elongated} are the triangular prism -- elongated triangular prism (ETDP), brick -- dipyramidal brick (DPB), and hexagonal prism -- elongated hexagonal dipyramid (EHDP)\footnote{A comparison of a triangular dipyramid to elongated triangular prism [not shown] with the same axial ratio demonstrates that the two dipyramidal shapes are nearly spectrally indistinguishable from one another}.  

As with the brick to non-brick comparison, the 8~--~40~\micron \ crystalline resonance features for each prism~--~dipyramid pair are strikingly similar to one another with the characteristics of a-, b-, and c-elongation/reduction clearly identifiable.  The positions and widths of the 10, 10.5, 16, and to a lesser degree, the 19~\micron \ features are the same, as expected from the elongation/reduction exercises on asymmetric bricks.  The largest differences between the prism~--~dipyramid shapes are the peak location of the 11~\micron \ feature, the width of the 11 and 23~\micron \ features, and the relative strengths of the 11, 19, and 23~\micron \ features. The dipyramidal shapes' spectral features generally have broader peaks and shoulders.  Specifically, in all cases except the 2~$\times$~b-elongated brick~--~b-elongated dipyramidal brick (DPB) comparison, the 11~\micron \ feature is weaker in relative strength with the presence of pyramidal tips, and in all cases it is shifted to shorter wavelengths.  The 19~\micron \ feature's behavior with the inclusion of pyramidal tips is dependent upon which of the three crystallographic axes is elongated.  For example, the 19~\micron \ feature systematically moves to shorter wavelengths and decreases in strength for the b-axis elongation case, while the 19~\micron \ feature's secondary long-wavelength peak becomes a weaker shoulder for the c-axis elongation case.  For a-, b-, and c-axis elongations, with the presence of pyramidal tips, the 23~\micron \ feature always is at longer wavelengths for the pyramid compared to its dipyramidal pair.  

Comparing rectangular prisms to non-rectangular prisms and to dipyramids (Fig.~\ref{fig:brick2nonbrick}, \emph{top, second, and bottom}), the rectangular prisms (with faces parallel to crystallographic axes) have the 23~\micron \ feature terminating at the shortest wavelength.   Inclusion of non-parallel faces extends the 23~\micron \ feature to longer wavelengths than for parallel faces, and to the longest wavelengths for dipyramids.  While significant, this shape characteristic is not degenerate with the elongation/reduction and asymmetry grain shape characteristics.  Instead, it adds a layer of detail onto our understanding of what characteristics of grain shape determine the precise resonance feature shapes, peak positions, and relative strengths.  The differences between the spectral features in Figs.~\ref{fig:brick2nonbrick} and ~\ref{fig:2times_elongated} indicate that an important external shape characteristic is the presence of crystalline faces and edges that are not parallel to the crystallographic axes.  

%%%%%%%%%%%%%%%%%%%%%%%%%%%%%%%%%%%%%%%%%%%%%%%%%%%%
\subsection{Shape Classes: Columns, Platelets, and Equant} Ê% 4.5
\label{sec:shape_classes}

The previous sections demonstrated that the effect of grain shape on the 8~--~40~\micron \ spectral features can be primarily described in terms of elongation/reduction with secondary spectral feature modifications due to asymmetry (non-elongated/non-reduced axes having unequal lengths) and non-parallel faces including tips on dipyramidal shapes.   
Specifically, the spectral features primarily are determined by the precise crystallographic axial ratio with the columnar, platelet, and equant grains' spectral shapes being clearly identifiable with non-linearly superimposed characteristics of the second longest axis and subtle, yet potentially significant, alterations to the features coming from non-parallel \emph{vs.} parallel faces.  In \S~\ref{sec:results_axes_sym} (last paragraph) we detailed a suite of diagnostic features for each of the a-, b-, and c-axis elongations.  
In both the symmetric and asymmetric cases, the behavior of the spectral features with elongation/reduction typically exhibited a distinct change for `extreme' elongations ($X~\gtrsim~2.6$) and reductions ($X\lesssim~0.2$).  Using these suites of diagnostic features, we define seven shape classes:  a-, b-, and c-columns ($X \gtrsim~2.6~\times$); a-, b-, and c-platelets ($X \lesssim~0.2~\times$); and equant ($0.2~\lesssim~X~\lesssim~2.6$).  These shape classes serve as a basis set that can potentially be applied to astronomical observations to determine if the forsterite crystals are column-shaped (`whisker' or `ribbon') or platelet-shaped.  A practical application is using the shape classes to rule out `extreme' elongations/reductions, i.e., reject `whisker' or platelet shapes, and hence conclude equant shapes are present.  

\begin{figure}[!h]	%Fig. 9
ÊÊÊ \begin{center}
ÊÊÊ ÊÊÊ \epsfig{file=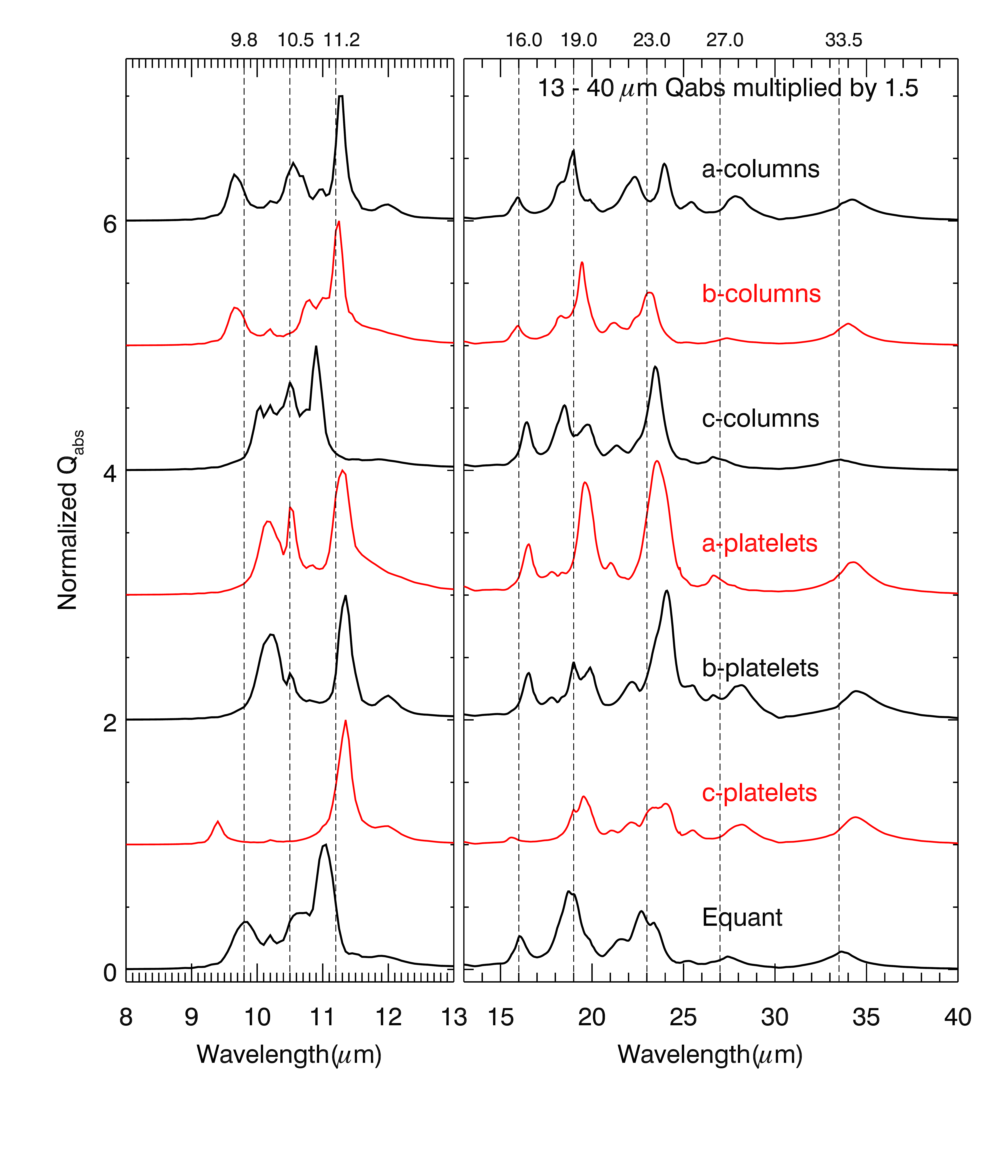,height=6.25in,width=6.5in}
        \caption{The 8~--~40~\micron \ normalized to the 11~\micron \ peak absorption efficiencies for the seven spectral shape classes defined in text.  The column shape classes are averages of all the symmetric and asymmetric bricks that are elongated by at least a factor of 2.6.  The platelet shape classes are averages of all the symmetric and asymmetric bricks that are reduced by at least a factor of 0.2.  The equant shape class is an average of all the symmetric and asymmetric bricks between elongation and reduction factors of 2.6 and 0.2.}ÊÊÊ ÊÊÊ 
ÊÊÊ \label{fig:shape_classes}
ÊÊÊ \end{center}
\end{figure}

The normalized absorptivities for the shape classes are shown in Fig~\ref{fig:shape_classes}.  Each shape class is an average of the symmetric and asymmetric bricks within the defined elongation/reduction bounds.  For each class, there are several 8~--~40~\micron \ diagnostic feature shapes, positions, and relative strengths that make them distinguishable from one another. The \emph{a-columns} have a 10.5~\micron \ feature that is equal to or stronger than the 10~\micron \ feature, a birfurcated 23~\micron \ feature, and an 11~\micron \ feature that is at least twice as strong as all the other features.  The \emph{b-columns} have singular peaks for all the features, an increased absorptivity tail on the long wavelength side of the 11~\micron \ feature, and a 19~\micron \ feature that is approximately twice as strong as the 23~\micron \ feature.  The \emph{b-columns} are most easily identifiable by determining that the spectral features are not consistent with the other shape classes. The \emph{c-columns} have a distinct 10~--~11~\micron \ feature complex that has a broad increased absorptivity shelf beginning near 10.0~\micron \ that is punctuated with distinct 10.5 and 11~\micron \ peaks; a bifurcated 19~\micron \ feature; and a singular 23~\micron \ feature that is significantly enhanced in strength.  The \emph{a-platelets} have a broad 10.2~\micron \ feature coupled with a narrow 10.5~\micron \ peak, and 19 and 23~\micron \ features that are nearly equal in strength and both enhanced in strength. The \emph{b-platelets} have a broad 10.2~\micron \ feature; lack a 10.5~\micron \ peak; have a 19~\micron \ feature with enhanced strength; and have a 23~\micron \ feature with significantly enhanced strength. The \emph{c-platelets}, of all the shape classes, have the largest separation between the 10 and 11~\micron \ peak positions with the 10 and 11~\micron \ features at 9.5, 11.35~\micron, respectively.  The \emph{c-platelets} have the 10~\micron \ feature reduced in strength to the point of being a weak feature, and the 19 and 23~\micron \ features are reduced and nearly equal in strength.

The \emph{equant} shape class in Fig.~\ref{fig:shape_classes} is the average of all symmetric and asymmetric bricks with a-, b-, and c-axis elongations and reductions spanning $0.2~\lesssim~X~\lesssim~2.6$ and with the non-altered axes having ratios of 0.8~:~1.2.  Constituents of the \emph{equant} shape class have crystallographic axes of nearly equal lengths so the features are sensitive to dominant effects of elongating/reducing the a-, b-, and c-axes and to the effects of the second longest axis.   For example, the a-axis can be the second or the third longest axis:  for c-elongated asymmetric bricks, the 10.5~\micron \ feature is in the 1.2 : 0.8 : 1.6 axial ratio brick but not in the 0.8 : 1.2 : 1.6 axial ratio brick, so the 10.5~\micron \ feature is present but is not dominant in the \emph{equant} shape class.  With the moderate elongations employed in the \emph{equant} shape class, the 10~\micron \ feature appears at 9.8~\micron \ and the 11~\micron \ feature appears at 11.0~\micron . The 19~\micron \ feature is a singular peak.  The 23~\micron \ feature is a rounded version of the 23~\micron \ trident feature, which is characteristic of the cube, as well as the other equant shapes of sphere and octahedron.  
The \emph{equant} shape class spectral features in descending order of strength are: 11, 19, 23, 9.8, 16, 27, and 33~\micron.    If spectral features are \emph{not} attributable to any of the six extreme elongations and reductions (columns and platelets), then probably the \emph{equant} shape class is relevant.   Spectra of comets and protoplanetary disks are strikingly similar to the \emph{equant} shape class (\S~{\ref{sec:disc_comets_disks}).

The strength of the 10.5 um feature alone should not be used as diagnostic for the equant class.  For equant shapes that are slightly elongated in the a-axis, either for the longest (Figs~\ref{fig:axial_ratios_8_13} and ~\ref{fig:axial_ratios_13_40}) or second longest (Fig.~\ref{fig:asymmetric_bricks_a}) axis, the 10.5~\micron \ feature is enhanced.  Even slight elongation ($X = 1.2$) of the a-axis produces a singular 10.5~\micron \ peak.  Alternately, an a-column identification can be made from a clear 10.5~\micron \ peak by examining the wavelength spacing between the 10 and 11~\micron \ peaks and the presence of a strong shoulder on the short wavelength end of bifurcated 23~\micron \ feature.
The remainder of the diagnostic features of the \emph{equant} shape class, however, remain clearly discernible.  The \emph{equant} shape class 23~\micron \ feature is an excellent representation of the trident-shaped feature of the sphere, cube and octahedron triple-peak and a good generalization of the width and shape of the tetrahedron, albeit the tetrahedron is double humped.  Generally, a-columns have a clearly bifurcated 23~\micron \ feature while equants are not bifurcated.

%%%%%%%%%%%%%%%%%%%%%%%%%%%%%%%%%%%%%%%%%%%%%%%%%%%%
\subsection{Effects of Grain Size on Resonant Features} Ê% 4.5
\label{sec:gsize_effects}

Grain size is an important factor when considering the spectral signature of forsterite grains in a natural setting.  Within the Rayleigh domain, where the grains are small compared to the wavelength of light, the infrared features are strongly influenced by the resonances in the bulk refractive indices and are frequently referred to as `surface modes' \citep{Bohren:1983,Min:2009CDNF}.   As the wavelength of light approaches the grain size, or the `size parameter' $x = 2\pi \times a_{\rm eff}/\lambda$ approaches unity, there is a phase lag of the wave inside the grain because the wavelength of the radiation inside the grain is small; the wavelength in the grain is small because the refractive index at the strong spectral features can be very high.  A feature becomes influenced by increasing 
grain size for increasing values of the product $m \times x = 2\pi \times (m/\lambda)$ \citep{Min:2005AA,Min:2009CDNF}, so shorter wavelength features (e.g., the 8~--~13~\micron \ features) are affected by increasing grain size at smaller grain sizes than the longer wavelength features (e.g., the 13~--~40~\micron \ features).  A good example is the comparison between DDSCAT computations of ellipsoids and CDE, which is a time-independent solution applicable when the phase lag in the grain is negligible \citep{Bohren:1983}: $Q_{\rm abs}$ agrees between DDSCAT and CDE for ellipsoidally shaped grains of sizes smaller than $a_{\rm eff}\simeq 0.5$~\micron, but for larger grains, the shapes of spectral features change with increasing grain size.  Grain sizes in comets and laboratory samples typically span submicron to micron-sizes, so grain size effects are best explored with the DDA method.  

In this section, the effect of grain size on the 8~--~40~\micron \ spectral features is investigated using DDSCAT for various grain shapes and sizes ranging from 0.1~--~3.0~\micron \ in effective radius.  In order to accomplish this goal, tetrahedra and elongated hexagonal dipyramids (EHDP) are computed with DDSCAT for ten grain sizes spanning effective radii Ê$a_{\rm eff}$ = 0.1~--~1.0~\micron\ with a linear step-size of 0.1 \micron \  (Fig.~\ref{fig:gsize_trends}).  The absorption efficiencies are normalized to the 11~\micron \ peak, the strongest peak for this range of grain sizes, to emphasize the change in the relative strengths of the features with changing grain size.  The general trend in the 8~--~13~\micron\ wavelength region is the peak positions of the spectral features shift to longer wavelengths as grain size increases, regardless of the external crystal shape.  For increasing grain size there is also a decrease in the relative strength of the 10 \micron \ feature, and in increase in relative strength for the 10.2 and 11.9~\micron \ minor peaks.  Understanding this behavior is essential to interpreting remote observations of comets since the 8~--~13~\micron\ region of a comet's spectrum is the most commonly observed and utilized spectral region for diagnosing the silicate mineralogy \citep{Hanner:1984,Wooden:1999}.  In the 13~--!40 \micron \ region, features do not exhibit significant shifts in peak wavelength position or feature shape, but do show a consistent increase in relative feature strength and broadening with increasing grain size.  

In Fig.~\ref{fig:gsize_trends}, when the grain size increases from 0.1 to 1.0~\micron \ the relative strength doubles for all the 13~--~40~\micron \ features. 
As the grain size increases, the same trends in the 8~--~13 \micron \ wavelength region seen for the tetrahedral grains are evident for the EHDP grains.  For both grain shapes, the locations of features at approximately 9.6~\micron\ (shoulder for tetrahedra and peak for EHDP) and 11~\micron \ shift to longer wavelengths.  
The amount of the shift of the 11~\micron \ peak towards longer wavelengths is less, however, for EHDP ($\sim$ 0.1 \micron ) than for tetrahedra ($\sim$ 0.3 \micron).  
The largest differences in feature shape with increasing grain size are that EHDP grains lose a short wavelength side shoulder on the 11~\micron \ feature, and the largest change for the tetrahedra grains are the rounding out of the 10~\micron \ peak and the 23~\micron \ feature changing from a double-peak to a feature, which is better described as a short wavelength side shoulder with an increasingly strong long wavelength side peak.  The difference in the amount of peak position shift between the two grain shapes indicates that grain size and grain shape effects are intertwined.  The degree to which the peak positions shift with increasing grain size depends upon the shape characteristics of the grain.  Nevertheless, the direction of the peak position shifts with increasing grain size remains the same regardless of specific grain shape.

\begin{figure}[!h] Ê %Figure 10
ÊÊÊ \begin{center}
ÊÊÊ ÊÊÊ \epsfig{file=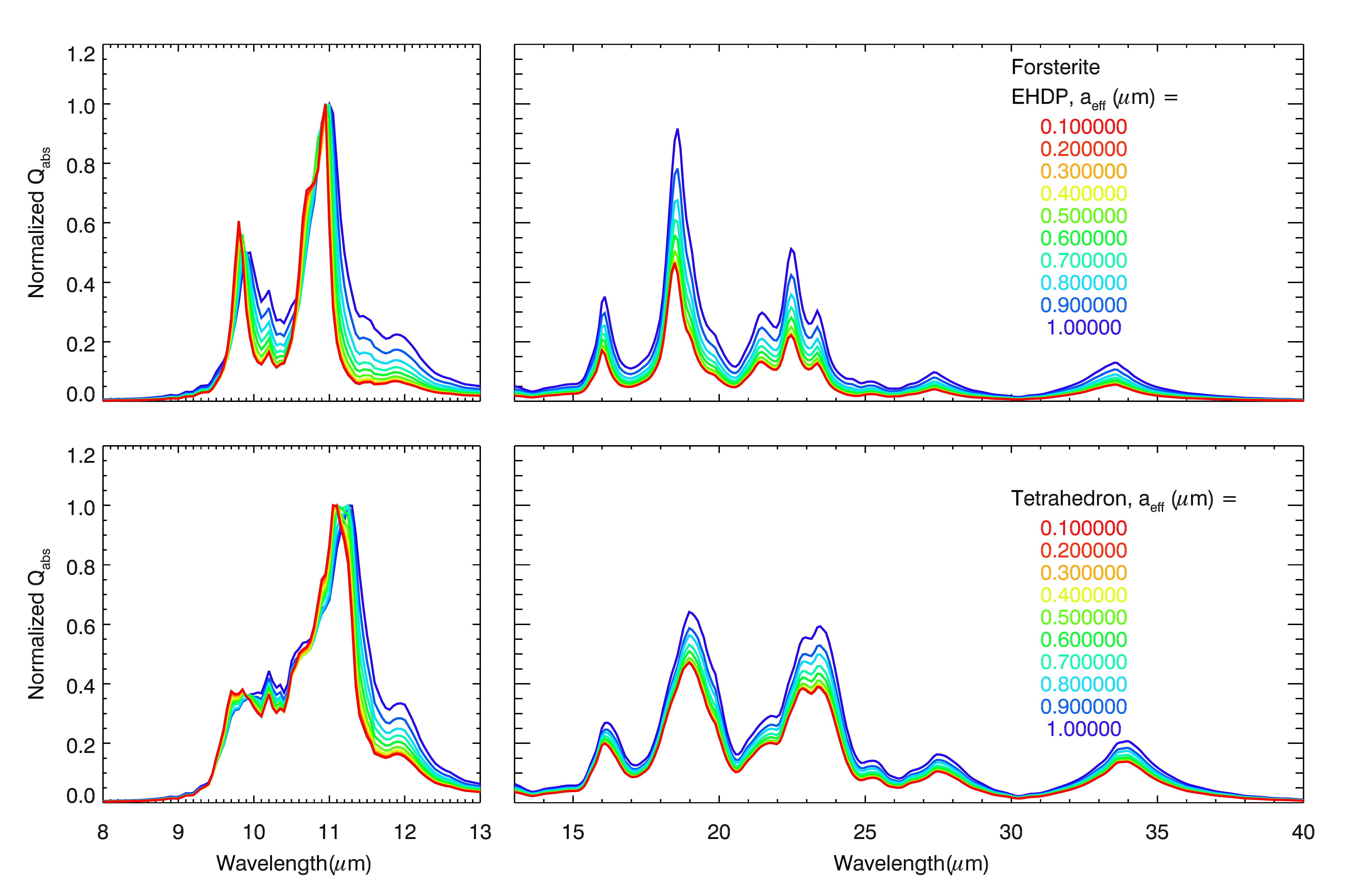, height=4.0in, width=6.0in}
ÊÊÊ ÊÊÊ \caption{The normalized to the 11~\micron \ peak absorption efficiencies from 8~--~13~\micron \ (\emph{left panels}) and 13~--~40~\micron \ (\emph{right panels}) for DDA computed elongated hexagonal dipyramids (EHDPs; \emph{top panels}) and tetrahedral (\emph{bottom panels}) forsterite. ÊThe effective radius is varied from 0.1~\micron\ to 1.0~\micron\ over 10 grain sizes linearly distributed with a 0.1 micron step--size.}
ÊÊÊ \label{fig:gsize_trends}
ÊÊÊ \end{center}
\end{figure}

For forsterite, as grain size increases above $\sim$0.4~\micron \ we see significant changes in the 8~--~40 \micron \ spectral shape due to the resonant features no longer being within the Rayleigh domain.  Outside the Rayleigh domain, spectral features change their spectral feature shape, peak location, and relative strength depending on the wavelengths and the strength of the each of the resonances.  Understanding these size dependency effects provides a diagnostic to determining the presence or absence of large grains in a sample of forsterite.  

In the previous examples, the grain sizes were limited to $a_{\rm eff} \le$~1.0~\micron \ in size to clearly show the trends in feature location and strength with increasing grain size.  Here, we extend the grain sizes to $\ge$1~\micron \ to show the dramatic changes in feature shape that occur when well outside the Rayleigh domain.  Figure~\ref{fig:gsize_large} displays the grain shape of a forsterite dimension (FD) brick  (a~:~b~:~c = 0.47~:~1~:~0.59, the axial ratio of forsterite's unit cell) for grain sizes $0.1 \le a_{\rm eff} \le 3.0$~\micron \ with a grain size resolution of $\Delta a_{\rm eff}$ = 0.04~\micron \ for $0.1~\le~a_{\rm eff}~\le~1.0$~\micron \ and $\Delta a_{\rm eff}$ = 0.4~\micron \ for $1.0~<~a_{\rm eff}~\le~3.0$~\micron.  The \emph{bottom} spectra in Fig.~\ref{fig:gsize_large} show all of the $Q_{\rm abs}$ over-plotted and unnormalized to emphasize the changes to feature shape, peak location, and \emph{absolute} feature strengths.  The \emph{top} spectra in Fig.~\ref{fig:gsize_large} singles out $a_{\rm eff}$ = 0.1, 1.6, and 3.0~\micron \ to more clearly elucidate large grain size spectral effects to the crystalline resonances.  The un-normalized absorption spectra confirm the expectation that the shorter wavelength spectral features leave the Rayleigh domain at smaller grain sizes than the longer wavelength spectral features.   Comparing the $a_{\rm eff}$ = 0.1, 1.6, and 3.0~\micron \ isolated curves, shows that by $a_{\rm eff}$ = 1.6~\micron, the 8~--~13~\micron \ features have significantly changed in their absorptive behavior, while the 13~--~40~\micron \ features have primarily only increased in strength and remained similar in shape and wavelength location. At a grain size of $a_{\rm eff}$ = 3.0~\micron, all but the 27 and 33~\micron \ spectral features have left the Rayleigh domain, and the features have begun to significantly change in shape, peak location, and strength.  In general, when a feature leaves the Rayleigh domain, the narrow crystalline resonances broaden significantly, become rounded, and decrease in peak absorption.  Since the 11~\micron \ feature diminishes in strength at smaller grain sizes than the 19 and 23~\micron \ features, at intermediate grain sizes the strongest spectral feature changes from the 11~\micron \ peak to either the 19 or 23~\micron \ peak.  Whether the 19 or 23~\micron \ peak is the strongest feature is dependent on the specific grain shape and grain size.  In a power-law type grain size distribution, the smallest grains will dominate the surface area while the larger grains will dominate the mass.  Despite the predominance by number of small grains, the presence of grains larger than $\sim$1~\micron \ can be discerned in a grain size distribution.  In astrophysical objects containing forsterite, e.g., the comae of comets, whether the 11 or 19/23~\micron \ feature is the strongest feature in the crystalline spectral component can serve as an observational spectral diagnostic to indicate whether or not a significant number of larger ($\geq$1~\micron ) grains are present.    

\begin{figure}[!p] Ê %Figure 11
ÊÊÊ \begin{center}
ÊÊÊ ÊÊÊ \epsfig{file=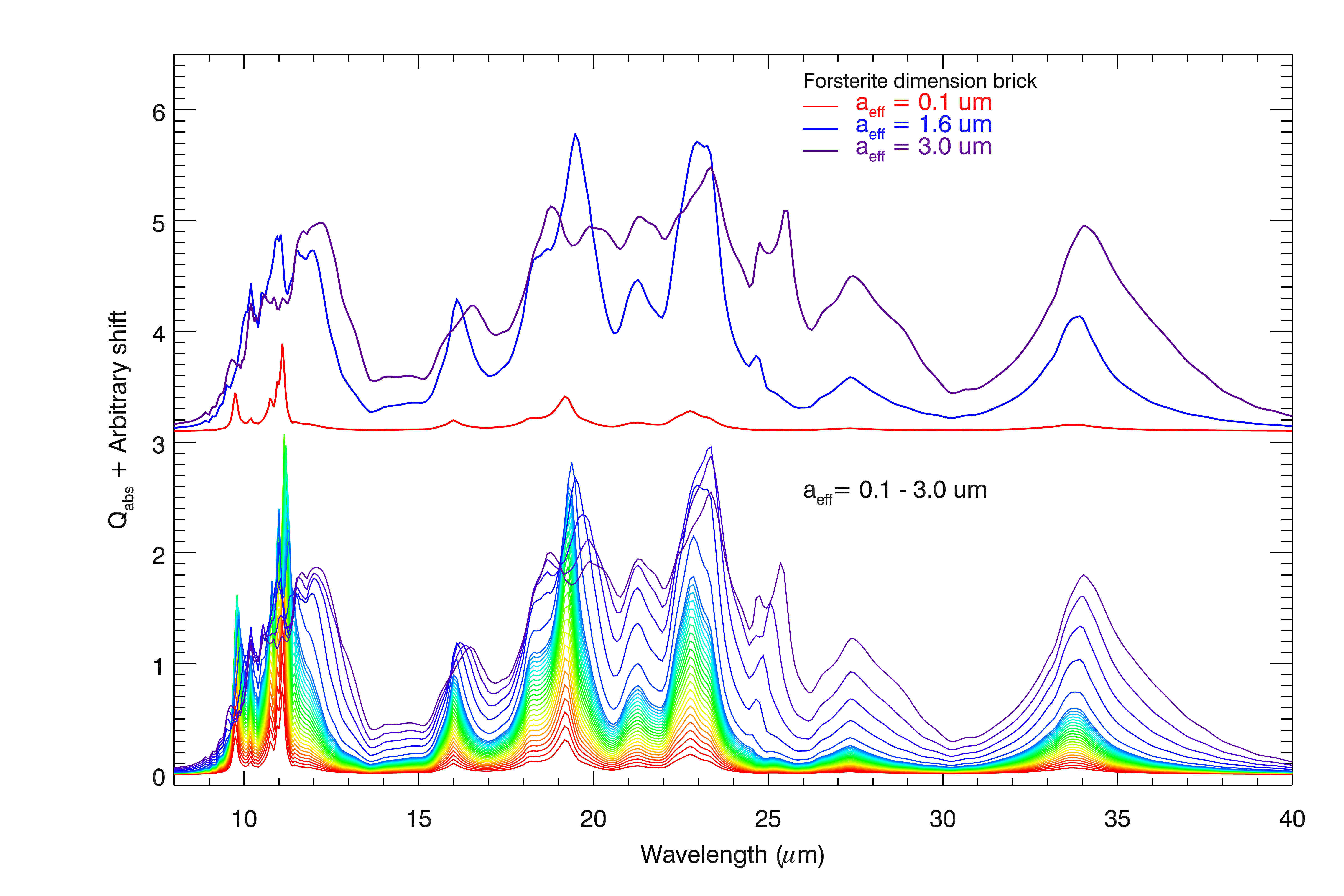, height=4.0in, width=6.0in}
ÊÊÊ ÊÊÊ \caption{The 8~--~40~\micron \ absorption efficiencies for DDA computed forsterite dimension bricks (asymmetric bricks with a~:~b~:~c = 0.47~:~1~:~0.59) forsterite grains for grain sizes $a_{\rm eff}$ = 0.1~--~3.0~\micron \ with a linear step size of 0.04~\micron \ for $a_{\rm eff}$ = 0.1~--~1.0~\micron \ and 0.4~\micron \ for $a_{\rm eff}$ = 1.0~--~3.0~\micron \ (\emph{bottom spectra}). ÊThe \emph{top spectra} isolate the grain sizes, $a_{\rm eff}$ = 0.1, 1.6, and 3.0~\micron \ to clearly demonstrate the effects of grain size as the spectral features leave the Rayleigh domain.}
ÊÊÊ \label{fig:gsize_large}
ÊÊÊ \end{center}
\end{figure}

%%%%%%%%%%%%%%%%%%%%%%%%%%%%%%%%%
\subsection{Ensembles of Grain Shapes or Sizes}
\label{sec:results_ensembles}

When compared to the absorption characteristics of laboratory forsterite samples, our small, $a_{\rm eff}$~=~0.1~\micron \ grains have much narrower spectral features.  The discrepancy is due to multiple factors, some of which are inherent differences between physical samples and our discrete models of defined shape and size target grains, and others of which are inherent to the nature of laboratory samples.  In contrast to the single-size and single-shape DDA results we have presented thus far, any physical sample of forsterite contains both a distribution of grain shapes and a distribution of grain sizes.  Additionally, laboratory prepared samples have grain properties that are related to the sample preparation techniques and the sample measurement techniques.  Laboratory samples are ground down by either hand or ball-grinding techniques, which produce characteristic grain sizes and shapes, and induce crystalline lattice distortions, all of which have been demonstrated to affect the shape and width of spectral features \citep{Koike:2010, Tamanai:2006ApJ, Tamanai:2009CDNF, Imai:2009}.  Moreover, grinding introduces grain shapes, sizes, and crystal dislocations \citep{Imai:2009} that are not necessarily representative of astrophysical samples of forsterite. The difficulty in comparing model and laboratory results is further compounded by the measurement techniques.  In order to measure the mass absorption coefficients for the ground samples, the laboratory samples either have to dispersed in an aerosol \citep{Tamanai:2006ApJ,Tamanai:2009CDNF} or embedded in a `transparent' medium of Cesium Iodide (CsI), Potassium Bromide (KBr), or polyethylene (PE) \citep{Koike:2003,Koike:2006,Koike:2010}.  The aerosol technique limits the grain sizes ($a_{\rm eff} \le 1.0$~\micron) as they pass through an impactor that attempts to separate agglomerated grains.  Also, in the aerosol measurement chamber, the grains likely are electrostatically clumped.  Both of these factors have been shown to affect the shape, peak position, and relative strength of spectral features \citep{Tamanai:2006ApJ,Tamanai:2009CDNF}.  Further, the embedding a sample in a medium for measurement has been shown to systematically shift the resonant features to longer wavelengths but where the amount of the wavelength shift varies depending on the feature  \citep{Tamanai:2006ApJ}.

Spectra of laboratory samples have broader and smoother spectral features than any single DDSCAT target grain of $a_{\rm eff} = 0.1$.  
The spectral feature differences between laboratory samples and DDA modeled grains significantly lessens when we consider either an average of many brick shapes with different axial ratios (or shape ensemble, Fig.~\ref{fig:shape_ensemble}) or a single shape averaged and weighted over a grain size distribution (GSD) (or size ensemble, Fig.~\ref{fig:gsd_fit_to_lab}).  The laboratory data that is lofted in air is the Alfa Irregular (Alfa Irr.) from \citet{Tamanai:2006ApJ}, and the grain shapes in the sample are described as `irregular'.  The DDA modeled grains are an ensemble of bricks composed of all $a_{\rm eff} = 0.1$~\micron \ grains, and are averaged over elongations/reductions without preference to any crystallographic axis.  The two ensembles considered are: 1) an average of all the symmetric and asymmetric bricks (`all bricks ensemble') presented in \S~\ref{sec:results_axes_sym} and \ref{sec:results_axes_sym}; and 2) an average of all symmetric and asymmetric bricks where $0.4 \le X \le 2.0$ (`shape restricted ensemble').  The `shape restricted ensemble' is a brick-selected representation of the equant shape class described in \S~\ref{sec:shape_classes}.  The two ensembles are compared to a single a-axis reduced asymmetric brick (0.4~:~1.2~:~0.8) for reference.  Both the `all bricks ensemble' and the `shape restricted ensemble' have broader spectral features than the single brick shape.  This demonstrates that when a variety of grain shapes are considered, the spectral features are broader and more comparable to physical samples of forsterite. 

The width of the features for the shape ensembles are still not as wide as the laboratory samples, but this exercise only considers grain shapes and restricts grain size to $a_{\rm eff}$ = 0.1~\micron, which is a much smaller grain size than the average grains in the \citet{Tamanai:2006ApJ} sample.   When an ensemble of grain sizes or grain size distribution (GSD) is considered, the spectral features significantly broaden.  Fig.~\ref{fig:gsd_fit_to_lab} displays the \citet{Tamanai:2006ApJ} Alfa Irregular laboratory sample data compared to a Hanner GSD \citep{Hanner:1984} with minimum grain size, $a_0$ = 0.1~\micron; peak grain size, $a_{\rm p}$ = 0.4~\micron; maximum grain size, $a_{\rm max}$ = 1.0~\micron; and power-law slope, $N$ = 3.0.  All of the absorption efficiencies for the GSD are DDA computed absorption efficiencies for tetrahedrally-shaped grains.  These GSD parameters were chosen because they provide a realistic comparison to the \citet{Tamanai:2006ApJ} laboratory sample, which has unquantified grain size characteristics other than a maximum size limited by the aerosol injector at  $a_{\rm eff}$ = 1.0~\micron.  For comparison, the `shape restricted ensemble' is included in Fig.~\ref{fig:gsd_fit_to_lab} to exhibit the differences between the shape and size ensemble approaches.  With the GSD approach, the spectral features are broader than for the shape ensemble. In general, the GSD spectral features are more comparable to the laboratory sample.  The GSD approach also is able to better match the relative strengths of the laboratory spectral features than the shape ensemble approach for the single grain size of 0.1~\micron, as is expected from the results of our study on grain size effects \S~\ref{sec:gsize_effects}.

Neither the shape ensemble nor the GSD approach replicate the precise features of the laboratory features, but both approaches demonstrate that including the realistic characteristics of a variety of grain shapes and grain sizes is able to broaden the features and make the relative strengths of the features more comparable to what is observed in laboratory experiments.   It may be possible to provide high quality DDA-based model fits to laboratory data, but that is outside the goals of this investigation.  Further, the approaches presented here only account for grain shape and grain size and do not attempt to account for the other previously mentioned grain characteristics including grain agglomeration in aerosol samples that also have an effect on the spectral features.  Such an extended study probably would make the DDA-based models even more comparable to the laboratory data.

\begin{figure} ÊÊÊ	%Fig. 12
ÊÊÊ \begin{center}
ÊÊÊ ÊÊÊ \epsfig{file=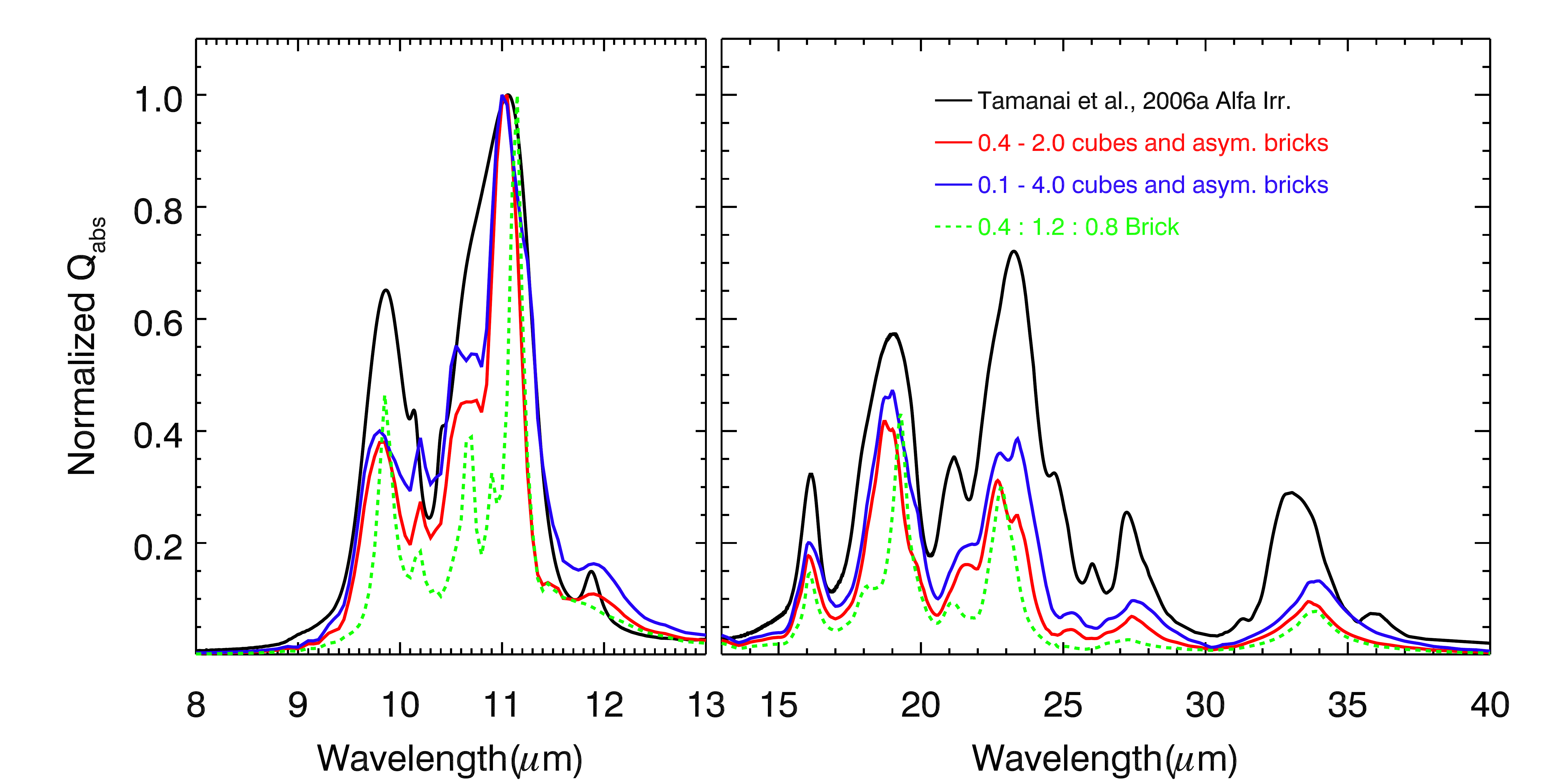,height=3.25in,width=6.5in}
        \caption{The normalized to the 11~\micron \ peak absorption efficiencies for a single 0.4~:~1.2~:~0.8 asymmetric brick (\emph{dashed green}), the `shape restricted ensemble' (\emph{solid red}) of all symmetric and asymmetric bricks where $0.4 \leq X \leq 2.0$, the `all bricks ensemble' (\emph{solid blue}) of all symmetric and asymmetric bricks, and the \citet{Tamanai:2006ApJ} aerosol technique measured Alfa Irregular laboratory sample (\emph{solid black}).  The widths of the features increases from the single shape to the `shape restricted ensemble' to the `all bricks ensemble'.}ÊÊÊ ÊÊÊ 
ÊÊÊ \label{fig:shape_ensemble}
ÊÊÊ \end{center}
\end{figure}

\begin{figure} ÊÊÊ	%Fig. 13
ÊÊÊ \begin{center}
ÊÊÊ ÊÊÊ \epsfig{file=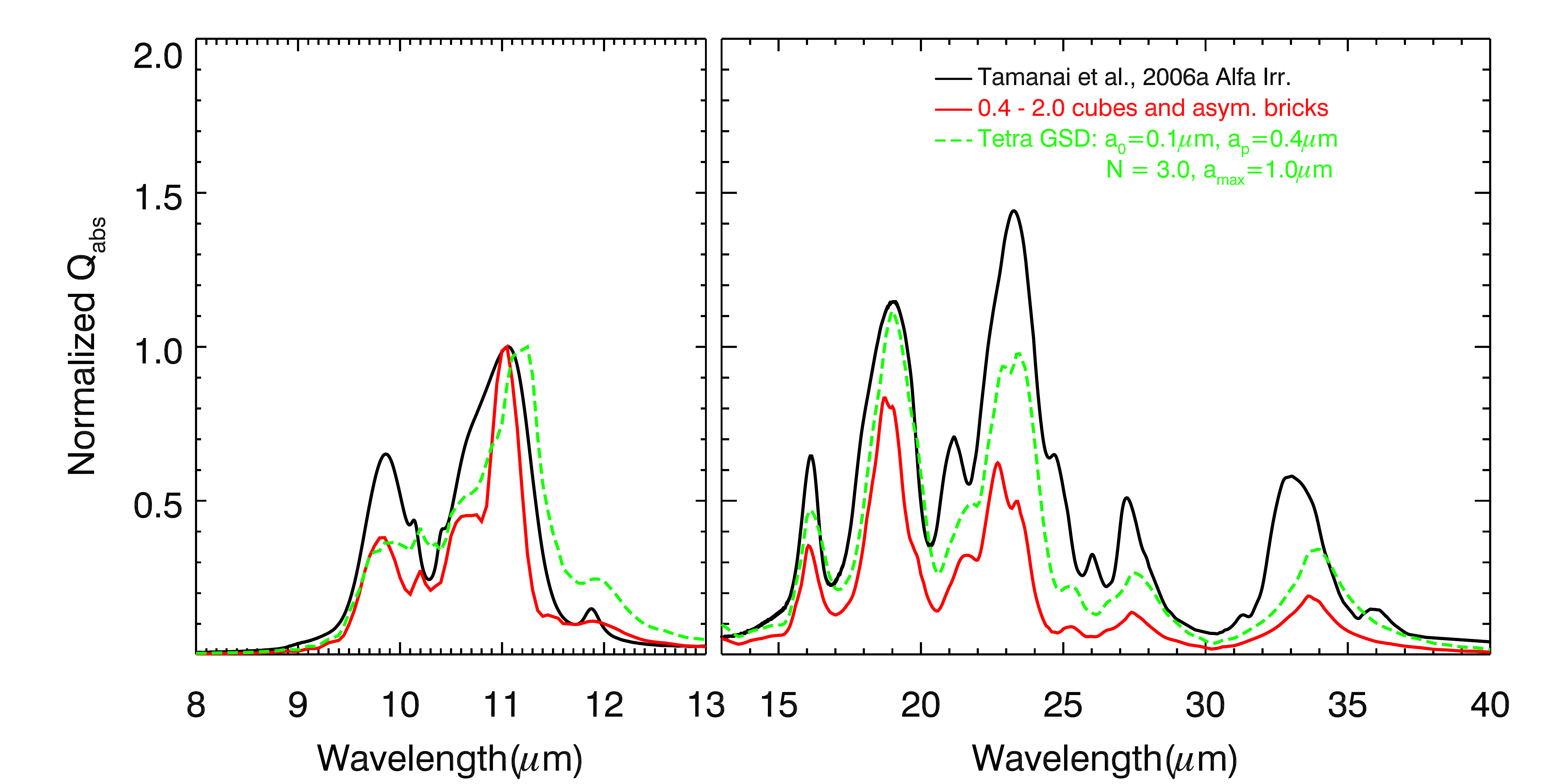,height=3.25in,width=6.5in}
        \caption{The normalized to the 11~\micron \ peak absorption efficiencies for a single 0.4~:~1.2~:~0.8 asymmetric brick (\emph{dashed green}), the `shape restricted ensemble' (\emph{solid red}) of all symmetric and asymmetric bricks where $0.4 \leq X \leq 2.0$; a Hanner grain size distribution (GSD) average with parameters $a_0$ = 0.1~\micron, $a_{\rm p}$ = 0.4~\micron, $a_{\rm max}$ = 1.0~\micron, and power law slope, $N$ = 3.0;  and the \citet{Tamanai:2006ApJ} aerosol technique measured Alfa Irregular laboratory sample (\emph{solid black}).}ÊÊÊ ÊÊÊ 
ÊÊÊ \label{fig:gsd_fit_to_lab}
ÊÊÊ \end{center}
\end{figure}

%%%%%%%%%%%%%%%%%%%%%%%%%%%%%%%%%%%%%%%%%%%%%%%%%%%%%%
\section{DISCUSSION}
\label{sec:disc}

%%%%%%%%%%%%%%%%%%%%%%%%%%%%%%%%%%%%%%%%%%%%%%%%%%%%%%
\subsection{Discrete Grains: Sensitivity to Grain Shapes with DDSCAT}
\label{sec:disc_sensitivities}

We summarize each of the spectral resonances sensitivity to grain shape characteristics (elongation/reduction of the longest crystallographic axis, asymmetry, and complex faces) in Table~\ref{tab:feature_sensitivities} and note that some features are more sensitive to certain shape characteristics than other features.  Certain spectral features exhibit varying degrees of overall sensitivity to grain shape such that some features are generally highly sensitive to all grain shape effects (10, 11, 19, 23, and 27~\micron \ features) or moderately sensitive (10.5, 16, 33.5~\micron \ features).  No single feature is  weakly-to-not sensitive to all investigated grain shape characteristics, but certain features are selectively sensitive to certain grain shape characteristics. 

The feature sensitivity designations are strongly, moderately, and weakly-to-not sensitive to a particular shape characteristic, where the boundary between moderately sensitive and strongly sensitive is defined by whether or not the shape characteristic changes the spectral feature's shape as well as peak wavelength location and strength.  Table~\ref{tab:feature_sensitivities} indicates that the 23~\micron \ feature is the only feature that is sensitive to all of the grain shape characteristics, accentuating this feature's importance in analysis of thermal IR spectra of astrophysical objects containing forsterite.  The 11~\micron \ feature, on the other hand, only shows strong sensitivity to the presence of (di)pyramidal tips.  The peak location of the 11~\micron \ feature, however, varies significantly with all of the shape characteristics, but it is only with the addition of dipyramidal tips that the 11~\micron \ feature exhibits a significant change in feature shape away from a well-defined singular peak.

\begin{table} \scriptsize \renewcommand{\arraystretch}{0.8} \vspace{0.005in}
\caption{Shape Characteristic Feature Sensitivities}
\begin{center}
\begin{tabular}{ p{3.5cm} p{0.9cm} p{0.9cm} p{0.9cm} p{0.9cm} p{0.9cm} p{0.9cm} p{0.9cm} p{0.9cm}} 

\toprule
Characteristic & 10$\mu$m & 10.5$\mu$m & 11$\mu$m  & 16$\mu$m & 19$\mu$m & 23$\mu$m & 27$\mu$m  & 33.5$\mu$m  \\ \midrule

a-elongation	& \ch& \ch& \ch&--& \ch& \ch& \ch& \ch \\ \midrule
b-elongation	& \ch	& X  & \ch	& -- & \ch &\dch	& -- & \ch \\ \midrule
c-elongation	& \dch & \dch &\ch &\ch &\dch &\dch	&\dch & -- \\ \midrule
a-reduction	& \dch & \dch & \ch & \ch & \dch & \dch & \dch & \ch \\ \midrule
b-reduction	& \dch & \dch & \ch & \ch & \dch & \dch & \ch & \ch \\ \midrule
c-reduction	& \ch & X & \ch & \ch & \dch & \dch & \ch & \ch \\ \midrule \midrule
elong./reduc.	& \dch & \dch & \ch & \ch & \dch & \dch & \dch & \ch \\ \midrule
Asymmetry	& \ch & \dch & \ch & \ch & \dch & \dch & \dch & \ch \\ \midrule
Pyramidal Tips	& \ch & \ch & \dch & -- & \ch & \dch & \ch & \ch \\  \bottomrule

\mbox{ --  Not-to-Weakly Sensitive: Little to no change in feature}
\mbox{\ch ~~Moderately Sensitive: Peak location changes and minor strength changes,i.e., no shape change}
\mbox{\dch ~~Strongly Sensitive: Feature shape, location, and strength changes}
\mbox{X:  Feature not present}

\end{tabular}
\label{tab:feature_sensitivities}
\end{center}
\end{table}

In addition to cataloging the general sensitivities of the 8~--~40~\micron \ spectral features, we select a sample of the DDA grain shapes that represent the grain shape characteristics detailed in \S~\ref{sec:results} (shown in Fig.~\ref{fig:the_figure}).  For this sample, we identify the wavelength location of the peaks and shoulders of the spectral features (summarized in Table~\ref{tab:the_table}).  The selected shapes are three columnar elongated triangular dipyramids (ETDPs) that are elongated in the a-, b-, and c-axes; three symmetric platelets that are reduced in the a-, b-, and c-axes; three equant asymmetric bricks [1.6~:~0.8~:~1.2], [0.8~:~1.6~:~1.2], and [1.2~:~0.8~:~1.6]; and four approximately 1~:~1~:~1 equant shapes: a tetrahedron, octahedron, cube, and sphere.  Such a selection includes elongated/reduced, asymmetric, and dipyramidal shapes, and a comparison to Fig.~\ref{fig:shape_classes} demonstrates that each shape is well represented by its shape class.  Table~\ref{tab:the_table} provides the quantitative basis and Fig.~\ref{fig:the_figure} provide a graphical representation to evaluate which spectral features are most sensitive to the effects of grain shape.  

\begin{figure} ÊÊÊ	%Fig. 14
ÊÊÊ \begin{center}
ÊÊÊ ÊÊÊ \epsfig{file=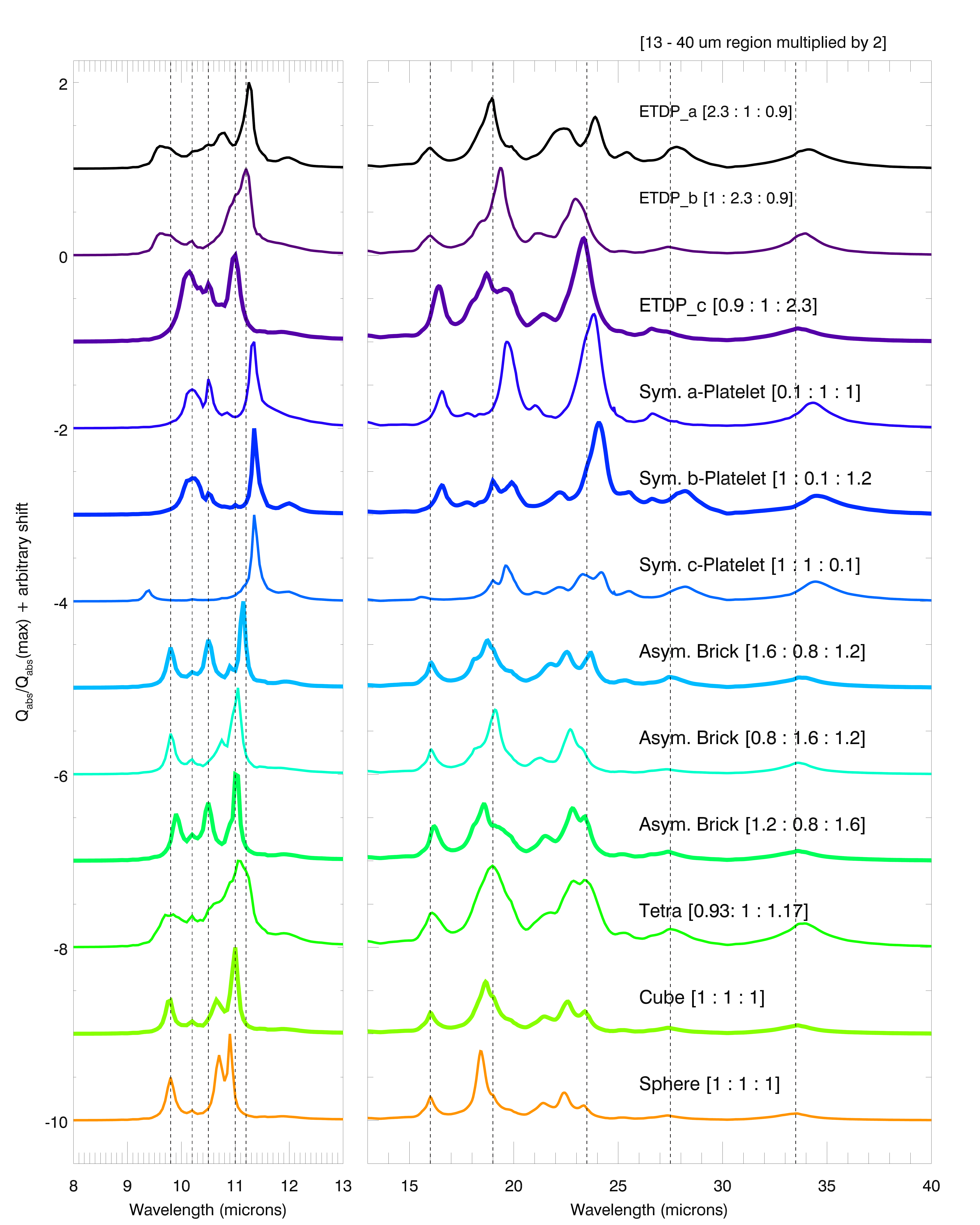,height=5.25in,width=6.5in}
        \caption{The normalized to the 11~\micron \ peak absorption efficiencies for a grain shape characteristic representative sample of the DDSCAT computations presented in this paper.  The grain shapes include three columnar shapes (ETDP$_a$, ETDP$_b$, and ETDP$_c$; Elongated Triangular Dipyramids), three platelet shapes (Symmetric a-, b-, and c-Platelets), three asymmetric equant shapes (asymmetric bricks), and three 1~:~1~:~1 equant shapes (tetrahedron, cube, and sphere). This set of grain shapes represents all seven shape classes and all three grain shape characteristics presented in this paper.  A summary of the wavelength positions of the spectra's peaks and shoulders is in Table~\ref{tab:the_table}.}ÊÊÊ ÊÊÊ 
ÊÊÊ \label{fig:the_figure}
ÊÊÊ \end{center}
\end{figure}

\begin{table}[!p] \scriptsize \renewcommand{\arraystretch}{0.8} \vspace{0.05in}
\caption{Resonance Peak Locations}
\begin{center}
\begin{tabular}{ p{3cm} p{0.9cm} p{0.9cm} p{0.9cm} p{0.9cm} p{0.9cm} p{0.9cm} p{0.9cm} p{0.9cm} p{0.9cm} p{0.9cm}} 

\toprule
Shape & 10$\mu$m & 10.5$\mu$m & 11$\mu$m  & 11.9$\mu$m & 16$\mu$m & 19$\mu$m & 23$\mu$m  & 25$\mu$m & 27$\mu$m & 33$\mu$m \\ \midrule

a-axis Elongated &  \textbf{9.60}  &  \emph{10.20}  &  10.80  &  12.00  &  \emph{15.60*} &  \textbf{18.95} &  22.40  &  25.40  &  \emph{27.45*}  &  [\emph{33.60*}, \\
Triangular & \emph{9.80} &10.50 & \textbf{11.25} & & \textbf{16.00} &  \emph{19.90} & \textbf{23.90} & & \textbf{27.80} & \textbf{34.20}$^\dagger$] \\
Dipyramid & & &  & & &   & & & \emph{28.20*}   \\  \midrule

b-axis Elongated  & \textbf{9.60} & \textbf{10.20} & \emph{10.80*} & \emph{11.85*} & \emph{15.60*} & \emph{18.40}$^\dagger$ & 21.15 & -- & \textbf{27.40} & [\emph{33.55*}, \\
Triangular & \emph{9.80} & & \textbf{11.20} & & \textbf{16.00} & \textbf{19.35} & \emph{23.35} & &  & \textbf{33.975}]  \\
Dipyramid & & & & & & & \textbf{22.95}  \\ \midrule

c-axis Elongated &-- &  \textbf{10.15} & \textbf{11.00} & 11.85 & \textbf{16.40} & \emph{17.95*} & \emph{21.45} & \emph{25.20}  & \textbf{26.60} & \textbf{33.60} \\
Triangular & & 10.50  & 11.50 & &  & \textbf{18.70} & \textbf{23.35} & & \emph{27.35*} &  \\
Dipyramid & & &  & & & 19.55 &  & &  \\ \midrule

Brick & -- & \textbf{10.20}  & 10.85 & -- & \textbf{16.55}  & 17.80 & \emph{21.80*} & \emph{25.20} & \textbf{26.625} & \textbf{34.25$^\dagger$}  \\  
Symmetric &  & 10.50 & \textbf{11.35} &  &  & \emph{18.30} &\emph{23.40*} & & 27.80 & \\
a-platelet & & & & & &  \textbf{19.675} & 23.85 & & &  \\ 
& & & & & & 21.00 \\ \midrule

Brick & -- &  \textbf{10.20}$^\dagger$  &  \textbf{11.35}  & 12.00 &  \textbf{16.55}  & 17.75 & 22.20$^\dagger$ & \emph{25.60}  &  \emph{27.80*}  &  \textbf{34.50} \\ 
Symmetric  & & 10.50 & & &  & \emph{18.35*}  & \textbf{24.075} &26.625 &  \textbf{28.225} &  \\
b-Platelet & & & & & & \textbf{19.00} & & & &  \\ 
& & & & & & \textbf{19.90}  \\ \midrule

Brick &  \textbf{9.40}  &  10.20  & \textbf{11.35}  & 12.00 & 15.55 & 19.00 & 21.05$^\dagger$ &  25.525  & \textbf{28.225} &  \textbf{34.425} \\ 
Symmetric & & & & & & \textbf{19.60} & 22.175 & &  &  \\
c-Platelet & & & & & &\emph{19.90*} &  \textbf{23.30}  \\ 
& & & & & & & \textbf{24.20} \\ \midrule

Asymmetric  &  \textbf{9.80}  & 10.20  & \textbf{11.15}  &  \emph{11.95}  & \textbf{16.05} &  \emph{18.05} & 21.775 & \emph{25.00*} & \emph{26.45}   &  \textbf{[33.60,} \\
Brick & & \textbf{10.50} & \emph{11.50} & & &  \textbf{18.75} & \textbf{22.55} & \emph{25.325} & \textbf{27.45}& \textbf{34.00]} \\
1.6 : 0.8 : 1.2 & & & & & & \emph{19.00*} &  \textbf{23.71} & & \emph{27.80*} \\ \midrule

Asymmetric  &  \textbf{9.80}  & 10.20  & 10.75  &  \emph{11.85}  & \textbf{16.05} &  \emph{18.10} & 21.15$^\dagger$ & 25.15$^\dagger$ & \emph{26.50*}   &  [\textbf{33.625}, \\
Brick & & & \textbf{11.05} & \emph{12.60*} & &  \textbf{19.10} & \textbf{22.70} & & \textbf{27.35} & \emph{34.00}] \\
0.8 : 1.6 : 1.2 & & & \emph{11.50} & & & &  \emph{23.35} & &  \\ \midrule

Asymmetric  &  \textbf{9.90}  & 10.20  & \textbf{11.025}  & 11.90  & \textbf{16.20} &  \emph{18.05} & 21.525 & 25.00$^\dagger$ & \emph{26.55}   &  [\textbf{33.60,} \\
Brick & & \textbf{10.50} & \emph{11.50} & & &  \textbf{18.575} & \textbf{22.80} & \emph{25.325} & \textbf{27.40}& \emph{34.00}] \\
1.2 : 0.8 : 1.6 & & & & & & \emph{19.30} &  \textbf{23.40} & & \emph{27.80*} \\ \midrule

Tetrahedron & 9.70 & 10.20  &  \emph{10.90} & 11.95 &  \textbf{16.075}  &  \emph{18.75*} & 21.80 &  25.30 &  \emph{26.55*}  & [\emph{33.60*}, \\
& \emph{9.85*} & \emph{10.50*} &  \textbf{11.05} & \emph{12.60*} & &  \textbf{19.00}  &  \textbf{22.85} & &  \textbf{27.475}& \textbf{34.00}]  \\
& & & & & & \emph{19.90} & \textbf{23.40}  \\ \midrule

%Octahedron & \textbf{9.80} & 10.55  &  \textbf{11.00} & 11.95 &  \textbf{16.075}  &  \emph{18.75*} & 21.80 &  25.30 &  \emph{26.55}  & [\emph{33.60}, \\
%& & &  \textbf{11.05} & \emph{11.50} & &  \textbf{19.00}  &  \textbf{22.85} & & \textbf{27.475} & \textbf{34.00}] \\
%& & & & & & \emph{19.90} &  \textbf{23.40}  \\ \midrule

Cube & \textbf{9.80} & 10.20  &  \textbf{11.00} & 11.90 &  \textbf{16.00}  &  \textbf{18.65} & \emph{21.475} &  25.15$^\dagger$ &  \emph{26.45*}  & \textbf{33.60} \\
& & 10.65 &  \emph{11.50} & \emph{12.60*} & &  \emph{19.05}  &  \textbf{22.575} & & \textbf{27.375} & \emph{34.00*}  \\
& & & & & & \emph{19.90} &  \textbf{23.40}  \\ \midrule

Sphere & \textbf{9.80} & 10.20  &  \textbf{10.90} & \emph{11.90} &  \textbf{16.00}  &  \textbf{18.40} & \textbf{21.40} &  \emph{25.15}$^\dagger$ &  \emph{26.45*}  & \textbf{33.55} \\
& & \textbf{10.70} &  \emph{11.50} & \emph{12.10*} & &  \emph{19.00}  &  \textbf{22.425} & & \textbf{27.35} &  \\
& & & & & & \emph{19.90} &  \textbf{23.35}  \\ \midrule

%$\Delta\lambda$(\micron) \ & 0.40 & -- & 0.35 & 0.15 & 0.55 & 1.275 & 1.775 & -- & 1.20 & 0.65 \\

\mbox{\textbf{Notation}: {\bf major peak}, peak, {\it shoulder}, \emph{* - minor shoulder}, $\dagger$ - Broad Peak }
\mbox{~~~~~~~ [~,~] - Flat-topped, two component peak.}

\end{tabular}
\label{tab:the_table}
\end{center}
\end{table}

Wavelength identification of the peaks and shoulders using an empirical algorithm (see Appendix) gives a numerical description of the features shape through the location of the peaks and shoulders and quantitatively highlights the degree of variation in feature location caused by grain shape.  The variation in the presence/absence of shoulders and minor peaks significantly varies over the presented grain shapes, highlighting the importance of grain shape in understanding the precise spectral structure of observations.  Without a proper treatment of the shape characteristics of forsterite in observations of comet comae and other astrophysical objects with crystalline forsterite spectral features, there is potentially a large problem in determining the precise mineralogy.  Since forsterite is the strongest crystalline silicate feature in comet comae, inaccurate modeling of the forsterite spectral features potentially may lead to the necessity of additional minerals to account for the spectral differences between the forsterite model and the data.

%%%%%%%%%%%%%%%%%%%%%%%%%%%%%%%%%%%%%%%%%%%%%%%%%%%
\subsubsection{Summarizing Resonant Feature Sensitivity to Grain Shape}
\label{sec:disc_sensitivity_sub}

Table~\ref{tab:the_table} provides the quantitative basis and Figs.~\ref{fig:the_figure} provide a graphical representation to evaluate which spectral features are most sensitive to the effects of grain shape.  This section analyzes the sensitivity of spectral features to grain shape for grains of $a_{\rm eff} = 0.1$~\micron, with the results grouped according to the feature wavelengths.  As a compliment to this section, refer to Table~\ref{tab:feature_sensitivities} that summarizes the sensitivity of each spectral feature to the effects of the shape characteristics of elongation, reduction, asymmetry, and the incorporation of pyramidal tips.  

\emph{The 10~--~12~\micron \ complex} has three strong features (10, 10.5, and 11~\micron) and two weak features (10.2~\micron \ and 11.9~\micron).  The strong features exhibit a high degree of variation over the different grain shapes.  For shapes where the c-axis is a strong component (c-axis columns and a-/b-axis reduced platelets), the 10~\micron \ feature becomes blended with the minor 10.2~\micron \ peak and appears as a well-defined peak at $\approx$~10.2~\micron.  For grain shapes that have a triangular component (e.g., tetrahedra, triangular dipyramid, and elongated triangular dipyramid), the 10~\micron \ feature appears more as a broad shelf.  This is a ubiquitous effect over all shapes with a `triangular' shape, or rather, the shapes that have many acute angles between their edges and faces.

The presence of the 10.5~\micron \ feature is dependent on predominantly the a-axis, but also has a weaker manifestation due to the c-axis.  When the a-axis is the longest or second longest axis, a 10.5~\micron \ feature will be present.  However, when the a-axis is minimized, the 10.5~\micron \ feature may not be present at all.  With the a-axis ETDP, the presence of dipyramidal tips affects the shape of the 10.5~\micron \ feature.  In this case, the broadening of the 11~\micron \ feature due to pyramidal tips is extreme, and the 11~\micron \ feature connects with the 10.5~\micron \ feature turning the 10.5~\micron \ feature in to a shoulder. 

The 11~\micron \ feature is a singular peak for nearly all of the grain shapes explored.  The exception is the `triangular' shapes, which have much broader 11~\micron \ features that are asymmetric primarily due to shoulders on the short wavelength side of the peak.  The peak position of the 11~\micron \ feature, however, is very sensitive to grain elongation/reduction of a crystallographic axis.  Over the shapes presented in this study, the 11~\micron \ feature peaks anywhere between 10.9 and 11.35~\micron.  The elongation of the a- and b-axes is responsible for shifting the 11~\micron \ feature to longer wavelengths, while elongation of the c-axis is responsible for moving the 11~\micron \ feature to shorter wavelengths. 

\emph{The 16~\micron \ peak} is similar in shape across all of our grain shapes, but exhibits variation in peak position and strength due to c-axis contributions.

\emph{The 19~\micron \ feature} exhibits significant variation in feature shape, peak positions, and strength, and is strongly sensitive to the elongation/reduction and asymmetry shape characteristics.  The strong sensitivity to these two shape characteristics implies that the  exact shape, peak locations, and strength are determined by the crystallographic axis ratio.  Regardless of the exact crystallographic axis ratio, for columns and platelets, the 19~\micron \ feature maintains the general characteristics of the associated shape class.

\emph{The 23~\micron \ feature} is the most sensitive to grain shape.  It exhibits significant variation in feature shape, peak positions, and strength, and is strongly sensitive to the elongation/reduction, asymmetry, and non-crystallographically parallel faces shape characteristics.  Over all shapes  the 23~\micron \ feature can appear as a single, double, or triple peak with complicated shoulder behavior depending on the grain shape characteristics.  For highly symmetric 1~:~1~:~1 equant grains (i.e., sphere, cube, elongated hexagonal dipyramid), the 23~\micron \ feature appears at shorter wavelengths and tends toward appearing as a `trident'-shaped triple peak.  Breaking the high degree of symmetry for 1~:~1~:~1 equant shapes (e.g., tetrahedron), however, removes the `trident'-shape, broadens, and shifts the 23~\micron \ feature to longer wavelengths,  

\emph{The weak 25~\micron \ feature} exhibits little variation over all the grain shapes.  In cases when the c-axis is a strong component, or when the 23~\micron \ feature terminates at long wavelengths, the 25~\micron \ feature will appear as a shoulder rather than a minor peak.

\emph{The 27~\micron \ feature} exhibits only small variations in feature shape, and peak location for grain shapes that are not strongly influenced by the c-axis, i.e., c-columns or a-/b-platelets. In cases where the c-axis has a strong influence, the 27~\micron \ feature is strongly sensitive, and it appears as a distinctive `shark tooth' shape that rapidly rises to the peak and slowly declines in absorptivity toward the 33.5~\micron \ feature.

\emph{The 33.5~\micron \ feature} exhibits moderate sensitivity over the grain shape characteristics with the largest differences being in peak location and broadness of the peak.  In Table~\ref{tab:the_table}, shapes with a notable influence from the a- and b-axes have a 33.5~\micron \ feature that is indicated as having a flat-topped two component peak.  These 33.5~\micron \ features also appear at longer wavelengths than features with strong c-axis influence.

Out of all the 8~--~40~\micron \ forsterite spectral features, only the 16 and 33.5~\micron \ features are moderately sensitive to all of the shape characteristics, implying that the character of forsterite grains' shape is a factor that cannot be overlooked when analyzing the other 8~--~40~\micron \ spectral features.  Interestingly, the 33.5~\micron \ feature in laboratory data nearly ubiquitously exhibits two companion peaks or shoulders on the short and long wavelength sides of the main peak \citep{Tamanai:2009CDNF, Koike:2010}.  The observed 33.5~\micron \ features for the comae of comets C/1995 O1 (Hale-Bopp) \citep{Crovisier:1997} at a heliocentric distance of 2.8 AU and 17P/Holmes during 2007 outburst \citep{Reach:2010} also contain shoulders at $\sim$32.8 and $\sim$34.7~\micron .  The presence of these additional shoulders on the 33.5~\micron \ features in physical samples and lack of them in our grain shape characteristic analysis suggests that these features are due to a grain characteristic other than those explored in this extensive investigation into forsterite shape characteristics.  

%%%%%%%%%%%%%%%%%%%%%%%%%%%%%%%%%%%%%%%%%%%%%%%%%%%%%%
\subsection{Model Applications}
\label{sec:implications}

%%%%%%%%%%%%%%%%%%%%%%%%%%%%%%%%%%%%%%%%%%%%%%%%%%%%%%
\subsubsection{Crystal Formation Environments}
\label{sec:disc_formation}

Crystalline silicates in comets are the product of either gas phase condensation or annealing of amorphous grains.  Condensation or annealing could have occurred in the inner 1 AU of the disk midplane where temperatures were $\gtrsim$ 1000 K \citep{Harker:2002, Wooden:2005CPPD, Wooden:2007PPV, Hanner:2010}.  During the early high mass accretion phase ($\gtrsim 10^{-6} M_\odot$/yr) of the disk, the region hot enough to condense or anneal crystalline silicates could have extended out to 3~--~4 AU \citep{Chick:1997, Boss:1998, Bell:2000}.  There is still debate as to which of the two crystal forming processes was primarily responsible for the forsterite that is observed in the comae of comets.  The grain shape of the crystals, however, has the potential to be a discerning factor in determining if the forsterite formed via condensation or annealing.

Laboratory experiments that evaluate forsterite grain shape as the result of condensation and partial evaporation indicate that crystal shape and environmental conditions (temperature and super-saturation) where the crystals form are causally linked \citep{Tsuchiyama:1998min,Bradley:1983,Kobatake:2008,Takigawa:2009}.  One such laboratory condensation experiment is the  \citet{Kobatake:2008} experiment, where the investigators find that crystals that form from a rapidly cooled highly supersaturated silicate vapor 
on to metal substrates are characterized by `bulky', `platy', `columnar/needle', and droplet shapes for values of temperature and supersaturation, T and $\sigma$, of 1000--1450$^\circ$C and $<97$, 700--1000$^\circ$C and 97--161, 580--820$^\circ$C and 131--230, and $<$500$^\circ$C and $>230$, respectively \citep{Kobatake:2008}.  Their experimental columnar/needle shapes, which form by vapor-liquid-solid growth process at lower temperatures ($<$820$^\circ$C), are extended stacks of plates, where the extension is not correlated with an axial direction.  Columnar/needles may be extended in the c-axis or a-axis direction, and can change directions, switching or combining  a- and c-axis extensions, and appear off-kilter or a bit askew in shape.  Other laboratory experiments find similar shape results, but there is not an exact agreement between crystal shape and environmental temperature or other conditions \citep{Tsuchiyama:1998min, Takigawa:2009}.  Note that a preliminary report on condensation on to a forsterite single crystal substrate reveals rough spheroidal shapes \citep{Tsuchiyama:2012Kobe}, so the substrate composition and growth rate may affect crystal shape.  

While there is currently disagreement within the laboratory community between the forsterite crystal shapes produced and the environmental conditions, as the laboratory experiments become better and begin to converge, the seven spectral shape classes (\S~\ref{sec:shape_classes}) potentially can be applied to remote sensing observational data to determine the general grain shape characteristics present.  Then, using the laboratory experiment results we can connect those shape characteristics to crystal formation environmental conditions.  Such a link between spectral features, grain shape characteristics, and formation conditions is relevant to cometary comae, several of which have clear 8~--~40~\micron \ spectral features due to forsterite.  Since cometary materials have remained unaltered since their formation and incorporation into comets, the determination of the crystal shape characteristics potentially provides a probe to the original crystal formation environmental conditions.  

As defined by \citet{Kobatake:2008}, the `bulky', `platy', and `columnar/needle' shapes  translate to the equant, platelet, and columnar shape classes (\S~\ref{sec:shape_classes}).  Figure~\ref{fig:shape_classes} shows the IR absorption efficiencies for the seven DDSCAT analog shapes that can be compared to the shapes of the condensed grains in \citet{Kobatake:2008}.  
From the descriptions and SEM images in \citet{Kobatake:2008}, some of the bulky shapes are multi-faceted and others look like bricks with axial lengths similar to forsterite dimensions with a~$<$~b~$<$~c.  We associate the bulky shapes, which are the highest temperature condensates (1270~--~1670~K), with the equant shape class.  We associate the platy shapes, which are the second highest temperature (970~--~1270~K) condensates, with a-, b-, or c-platelets.  We associate the a- and c-columnar shapes, which are coolest liquid--vapor--solid condensates ( ), with a- and c-columns.  B-columnar shapes do not appear to be one of the condensed shapes.   A-platlets and a- and c-columns have the 10.5~\micron \ feature.  A-platelets and c-columns have their 10~\micron \ feature at 10.2~\micron .  A- and b-platelets and c-columns have a significantly enhanced 23~\micron \ feature (relative to other features).  C-columns have their 11~\micron \ feature shortward of 11.0~\micron .  Comparing the suite of diagnostic wavelength positions and relative strengths of the spectral features allows an assessment of whether equant, a- or c-columns, or platelets are prevalent in spectra of astrophysical sources.  Identification of shape classes in astrophysical souces provides a potential means to probe the temperatures at which forsterite formed.  

%%%%%%%%%%%%%%%%%%%%%%%%%%%%%%%%%%%%%%%%%%%%%%%%%%%%%%
\subsubsection{Comets and Protoplanetary Disks}
\label{sec:disc_comets_disks}

Comet spectral energy distributions commonly show 9.8 and 11.2~\micron \ peaks without a strong and distinguishable 10.5~\micron \ peak \citep{Hanner:2004cometsii}.   In comet Hale-Bopp's SED observed near perihelion, the 10.5~\micron \ peak appears just as strong as the 11.2~\micron \ peak but the crystalline contribution to the 10.5~\micron \ peak is actually significantly weaker compared to the 11.2~\micron \ peak because the 10.5~\micron \ peak lies on the top of the amorphous silicate feature whereas the 11.2~\micron \ peak lies on the trailing shoulder of the amorphous silicate feature \citep{Wooden:1999}.  Hence, the 10.5~\micron \ peak in comets is not distinguishable and not strong, suggesting that the grains do not belong to the a-column, c-column, and a-platelet shape classes (\S~\ref{sec:shape_classes}, Fig.~\ref{fig:shape_classes}).  

The cometary 11.2~\micron \ feature actually spans 11.05~--~11.2~\micron \ in high signal-to-noise spectra of comets Hale-Bopp  \citep{Crovisier:1997, Wooden:1999, Harker:2002,Harker:2004err}, 17P/Holmes after its outburst \citep{Reach:2010}, and 9P/Tempel~1 following Deep Impact \citep{Lisse:2006Sci}.  The 11.05~--~11.2~\micron \ wavelength range for the peak excludes c-column grains as a component of Hale-Bopp's forsterite because c-column grains have their 11~\micron \ feature short-ward of 11.0~\micron.  These same three comets have their 10~\micron \ feature spanning 9.8~--~10.1~\micron.  C-platelet grains have a very weak 10~\micron \ feature at approximately 9.6~\micron, so the comet grains likely do not belong to the c-platelet shape class.  A significantly enhanced 23~\micron \ feature relative to other features also is not observed for these comets so the grains likely are not c-columns or a-/b-reduced platelets.  B-columns do not have a significantly enhanced 23~\micron \ feature and do have their 11~\micron \ peaks near to 11.2~\micron, so b-columns are a likely spectral class in comets.  
The equant shape class produces the 11 and 23~\micron \ features at 11.0 \emph{and} 23.5~\micron \ and the strongest feature is at 23.5~\micron .  Specific axial ratios for equant shapes produce feature positions and relative strengths (Fig.~\ref{fig:the_figure}) that are in good agreement with the spectra of these three comets:  tetrahedra have their 11~\micron \ feature at 11.05~\micron \ and a double-humped 23~\micron \ feature with peaks at 22.85 and 23.40~\micron .  Comparison of spectra of three bright comets with the seven shape classes suggests that the shapes of forsterite crystals in the comae of comets are equant or preferentially moderately elongated along the b-axis\footnote{Forsterite dimensions is on the boundary between equant and b-columns.}. 
Exploration of the effects of crystal shape on spectral feature shape, peak positions, and relative strengths as well as exploration of the effects of grain size distributions sets the ground-work for fitting comet spectra with grain size distributions of equant or b-elongated forsterite grains and is within the scope of \citep{Lindsay:2013PII}. 

A brief examination of external protoplanetary disk spectra show similar trends in spectral features as with comets, which historically has been shown through the classic comparison of comet Hale-Bopp with HD100546 \citep[][Fig.~4,]{Malfait:1998}.  The 10.5~\micron \ peak is not distinguishable, and the 23~\micron \ peak is not overwhelmingly dominant \citep{Olofsson:2009,Oliveira:2011}.   Likely, the grain shapes revealed by comet and disk observations are equant, and thus the precise spectral features are dependent on the crystallographic axis ratio.  While this type of analysis cannot for certain determine the formation history of the crystalline grains, by the \citet{Kobatake:2008} condensation experiment, the analysis suggests that the environmental conditions under which the crystals formed favored condensation at temperatures greater than 1000$^\circ$C (1273 K).  Condensation origin for forsterite crystals is strongly supported by laboratory examination of forsterite crystals in cometary IDPs \citep{Bradley:1983,Molster:2003}, which are described as having equant and tabular crystal habit (Bradley:2011, personal communication), and in Stardust samples \citep{Brownlee:2006,Nakamura:2008,Tsuchiyama:2009}.   Hence, our shape classes can be used in an effort to ascertain the formation conditions of forsterite crystals that are observed in astronomical sources.

%%%%%%%%%%%%%%%%%%%%%%%%%%%%%%%%%%%%%%%%%%%%%%%%%%%%%%
\section{SUMMARY AND CONCLUSIONS}
\label{sec:summary}

Through our DDA computations using the publicly available code DDSCAT v.7.0 \citep{Draine:1994DDA,Draine:2008_70}, we identify three characteristics of grain shape that have distinct effects on the 8~--~40~\micron \ spectral features of forsterite with respect to feature shape, peak position, and strength.  In order of decreasing influence, the identified shape characteristics are: 1) the extent to which a crystal is \emph{elongated/reduced} along the longest/shortest crystallographic axis (\S~\ref{sec:results_axes_sym}); 2) the \emph{asymmetry} of the crystal shape such that all three crystallographic axes are of different length, i.e. the crystallographic axial ratio (\S~\ref{sec:results_axes_asym}); and 3) the presence of crystalline faces that are not parallel to a crystallographic axis, e.g., (di)pyramids and non-rectangular prisms (\S~\ref{sec:results_faces_edges}).  

The first shape characteristic exercise, elongation/reduction, demonstrated that under `extreme' elongation ($>~2.6~\times$ elongated along a single axis) and reduction ($<~0.2~\times$ reduced along a single crystallographic axis), the 8~--~40~\micron \ spectral features are distinct between the a-, b-, and c-axis elongations/reductions.  The discerning spectral features are primarily the shape, position, and strength of the 10, 10.5, 11, 19, 23, and 27~\micron \ features.  Using the distinctness of these spectral features between the elongation/reduction state, in \S~\ref{sec:shape_classes}, we define seven spectral shape classes: 1~--~3) a-, b-, and c-axis columns; 4~--~6) a-, b-, and c-axis platelets; and 7) equant.  Equant shapes are non-elongated/reduced to moderately elongated/reduced shapes that are nearly equal in all dimensions.  Each of spectral shape classes has a suite of spectral features that can be used to potentially positively identify, or rule out, the elongation/reduction shape characterisitic of forsterite.

The second shape characteristics exercise, asymmetry, demonstrates that the spectral characteristics of the second longest crystallographic axis are important to the spectral signatures of forsterite.  The spectral features for asymmetric bricks with three different axis lengths exhibit significant spectral differences between renditions of which crystallographic axis is the second longest axis.  Regardless of how long or short the primary crystallographic axis is, spectral characteristics related to the second longest axis are non-linearly superimposed on the general spectral shape class template.  The asymmetry effect particularly is apparent when the a-axis is the second longest axis.  In this case, a 10.5~\micron \ feature that is characteristic of a-axis elongation is present.  Such asymmetry effects can potentially lead to a misidentification of the spectral shape class, particularly for the equant shape class, which is the most sensitive to the effects of asymmetry due to the three axes being nearly equal in length.  However, an examination of all of the 8~--~40~\micron \ spectral features can prevent a misidentification.  

The third shape characteristic, crystallographically non-parallel faces, demonstrates that the alignment of crystalline faces with respect to the crystallographic axis orientation can have significant spectral effects.  This is accomplished by expanding the polyhedral shapes to include non-rectangular prisms and their (di)pyramidal pairs, which have the same axial ratios at the rectangular prisms (bricks).  The comparison between brick and non-brick shapes (Fig.~\ref{fig:brick2nonbrick}) confirm the expectation from the elongation/reduction and asymmetry results that shapes with similar axial ratios have similar spectral features.  The agreement between spectral features is strong enough that regardless of the shape differences, the non-brick shape spectral features are clearly recognizable as belonging to their respective spectral shape classes.  The main difference between the non-brick and brick shapes is the presence of crystal faces that are not parallel to the crystallographic axes implying that the spectral differences are due to the presence of crystallographically non-parallel faces.  The spectral effects of incorporating tips are subtle, and can be generally described as broadening the 10, 11, and 23~\micron \ features.  In summary, when the axial ratios are the same the differences between the spectral features are caused by the differences in grain shape.  The effect due to the shape characteristic of crystallographic non-parallel faces also is strongest with shapes that have a `triangular'-shape element, i.e., for tetrahedra, triangular dipyramids, and elongated triangular dipyramids.  

A quantitative comparison of the wavelength locations of spectral features for eleven DDSCAT polyhedral shapes and a sphere is given in Fig.~\ref{fig:the_figure} and in Table~\ref{tab:the_table}.  Sensitivities of the spectral features near 10, 10.5, 11, 16, 19, 23.5, 25, 27, and 33.5~\micron \ to grain shape are summarized in Table~\ref{tab:feature_sensitivities} and discussed in \S~\ref{sec:disc_sensitivities}.  When considering all shape characteristics, we recommend considering first the 10.5,
11 and 23.5~\micron \ features, then the 10, 19, 25, and 27~\micron \ features, and lastly 16 and 33.5~\micron \ features. 

A critical examination of the 8~--~40~\micron \ features with respect to grains size reveals that grain size alters the absorption resonances in distinguishable ways.  In a grain size distribution of a specific grain shape, as $a_{\rm eff}$ approaches 0.5~\micron \ from below , the 8~--~40~\micron \ spectral resonances begin to leave the Rayleigh domain.  At $a_{\rm eff} >$ 1.6~\micron, the changes in the shapes of the spectral features are no longer predictable based on the feature shapes of grains smaller than $\sim$1.6~\micron .  In general, as grain size increases, the features diminish in strength and significantly change their shape starting at the 10~--~11~\micron \ features and progressing to the longest wavelength features.  

The spectral shape classes defined in \S~\ref{sec:shape_classes} provide a practical means to analyze the shape characteristics of forsterite in comet comae observations.  Using the diagnostic spectral features related to a-, b-, and c-axis columns/platelets, it is potentially possible to positively identify the shape characteristics of the forsterite in comet comae.  By ruling out particular shape classes, the shape classes that possibly are present become apparent.  In \S~\ref{sec:disc_comets_disks}, we use the 10, 10.5, 11, and 23~\micron \ spectral features of those comet comae that have strong crystalline resonances to rule out the presence of platelet shapes and a- and c-axis columns.  By the process of elimination, cometary forsterite likely is equant or b-columnar in shape.  A more detailed analysis, however, is required to confirm this hypothesis.  

Based upon the results of condensation and partial evaporation experiments \citep{Tsuchiyama:1998min, Kobatake:2008, Takigawa:2009}, there is a causal link between the environmental conditions where forsterite forms and grain shape characteristics (\S~\ref{sec:disc_formation}).  By comparison with condensation experiments by \citep{Kobatake:2008}, equant grains may be high temperature condensates (1270~--~1670~K).  
Thus, knowledge of each of the grain shape characteristics can potentially provide insight into the formation environmental conditions of the Solar System's protoplanetery disk during the epochs of dust grain formation up to the time when these grains were incorporated into the small bodies of the Solar System.  Parallel implications can also be drawn for external protoplanetary disks containing significant amounts of forsterite.

%%%%%%%%%%%%%%%%%%%%%%%%%%%%%%%%%%%%%%%%%%%%%%%%%%%%%%

\section{ACKNOWLEDGEMENTS}
We express special appreciation for laboratory experimentalists Dr. Akemi Tamanai and Dr. Chioye Koike for sharing electronic versions of data published by the Jena and Kyoto groups, respectively.  We thank summer intern Brittany M. Hunter (University of Western Australia) for working with DHW on in-depth analyses of asymmetric bricks.  The authors wish to thank 
the efforts of an anonymous referee whose suggestions and critique greatly improved the manuscript.  
SSL thanks the Planetary Systems Branch of the NASA Ames Research Center for hosting him as a GSRP Fellow (Grant No.~NNX08AV43H/114) during his three 12 week GSRP Summer visits.  He also expresses his strongest gratitude to DHW for all the long hours and hard work she put in regards to this paper.  Her efforts are greatly appreciated!  SSL also thanks JRM for his guidance in assembling this paper.  
DHW, DEH and MSK acknowledge significant support from the Planetary Atmospheres (08-PATM08-0080) Program.
The DDSCAT computations utilized \emph{NAS} time allocations (SMD-09-1144, SMD-10-1637, and SMD-11-2361) were awarded to DHW by competitive process for the associated approved PATM research program.  SSL thanks CEW for is insightful edits and masterful word-smithing.
SSL and DHW would also like to thank Zack Gainsforth for his insights into the unit cell structure and crystal habit of forsterite.  We thank J.~Bradley, H.~Mutschke, Th.~Henning, and C.~J\"ager for comments during the progress of the analyses.  

\bibliographystyle{apj}
\bibliography{CometDust}

%%%%%%%%%%%%%%%%%%%%%%%%%%%%%%%%%%%%%%%%%%%%%%%%%%%%%%
\newpage
\appendix

\section{Algorithm for Identifying Peaks and Shoulders}  
\label{sec:appendix}
In order to consistently identify the peaks and shoulders in an absorption efficiency curve, we use an empirically derived algorithm that detects peaks and shoulders based on the curvature (second derivative) of the absorption efficiencies.  The classifications of peak or shoulder are based upon the  empirical $Q_{\rm abs}$ curvature function, which is defined by Eqs.~\ref{eqn:curvature_function_def}~and~\ref{eqn:qbar} where `sm' denotes the smooth function in the library of commands of the commonly used Interactive Data Language (IDL) \citep[e.g.,][]{Albrecht:1997IDL}.  If the strength of the curvature function is above a cut-off value for two or more consecutive wavelengths, then the related feature is identified by the algorithm and classified as a peak if there is a change in sign for the derivative of $Q_{\rm abs}$, and as a shoulder otherwise.  The threshold value is chosen such that the shoulder of the 27~\micron \ feature is included and less pronounced features are not identified.  When identifying a peak, if a neighboring pixel $Q_{\rm abs}$ value is within one percent of the maximum, the algorithm takes the average of the two wavelength positions as the peak location.  

There are cases when the curvature function has either a complex structure or is too narrow to return a proper identification of a peak or shoulder.  A complex ``M'' shape in the curvature function occurs when a feature is broad.  This is because the slope of the $Q_{\rm abs}$ is rapidly changing as it approaches the peak or shoulder, slowly changing across the broad peak or shoulder, and then again rapidly changing on the far side of the peak or shoulder creating the m-structure in the curvature.   For broad peaks, the algorithm selects the mid-wavelength position of the interval of the curvature function above the cut-off value for the feature in question.  For the weaker shoulders, a linear fit is made between the endpoints of the range of the curvature function above the cut-off value for the feature in question.  If the maximum of ratio of $Q_{\rm abs}$ to this line-fit is greater than 2\%, then the feature is identified as a weak shoulder.  This empirical method produces identifications consistent with identifications by eye and yields a consistent set of wavelengths for the peaks and shoulders.

\begin{center}
\begin{equation}
\label{eqn:curvature_function_def}
\rm{Curvature~Function} = -\bar{q}\times\frac{\delta^2\lambda}{\delta\lambda^2}\{\bar{q}\}
\end{equation}
\end{center}

\begin{center}
\begin{equation}
\label{eqn:qbar}
\bar{q} = { {{ \left\{ \frac{Q_{\rm abs}}{{\rm max}(Q_{\rm abs})}  \right\} ^2}} \over {{\rm sm}({\rm sm}({\rm sm}({\rm sm}({ \left\{ \frac{Q_{\rm abs}}{{\rm max}(Q_{\rm abs})}  \right\} } ,3),20),20),20)} }
\end{equation}
\end{center}

Fig.~\ref{fig:lo_method} depicts how the peak identification algorithm is implemented for three grain shapes: sphere, tetrahedron, and dipyramidal brick.  The DDSCAT  $Q_{\rm abs}$ values appear in black while the empirical curvature function is in red.  In this figure, the curvature function has been normalized and plotted with a logarithmic ordinate to illustrate how the peaks and shoulders are identified.  The letters above the features indicate the peaks (P), broad peaks (B),  shoulders (S), and broad shoulders (b).  

\begin{figure}[!p]  %Figure 20
    \begin{center}
        \epsfig{file=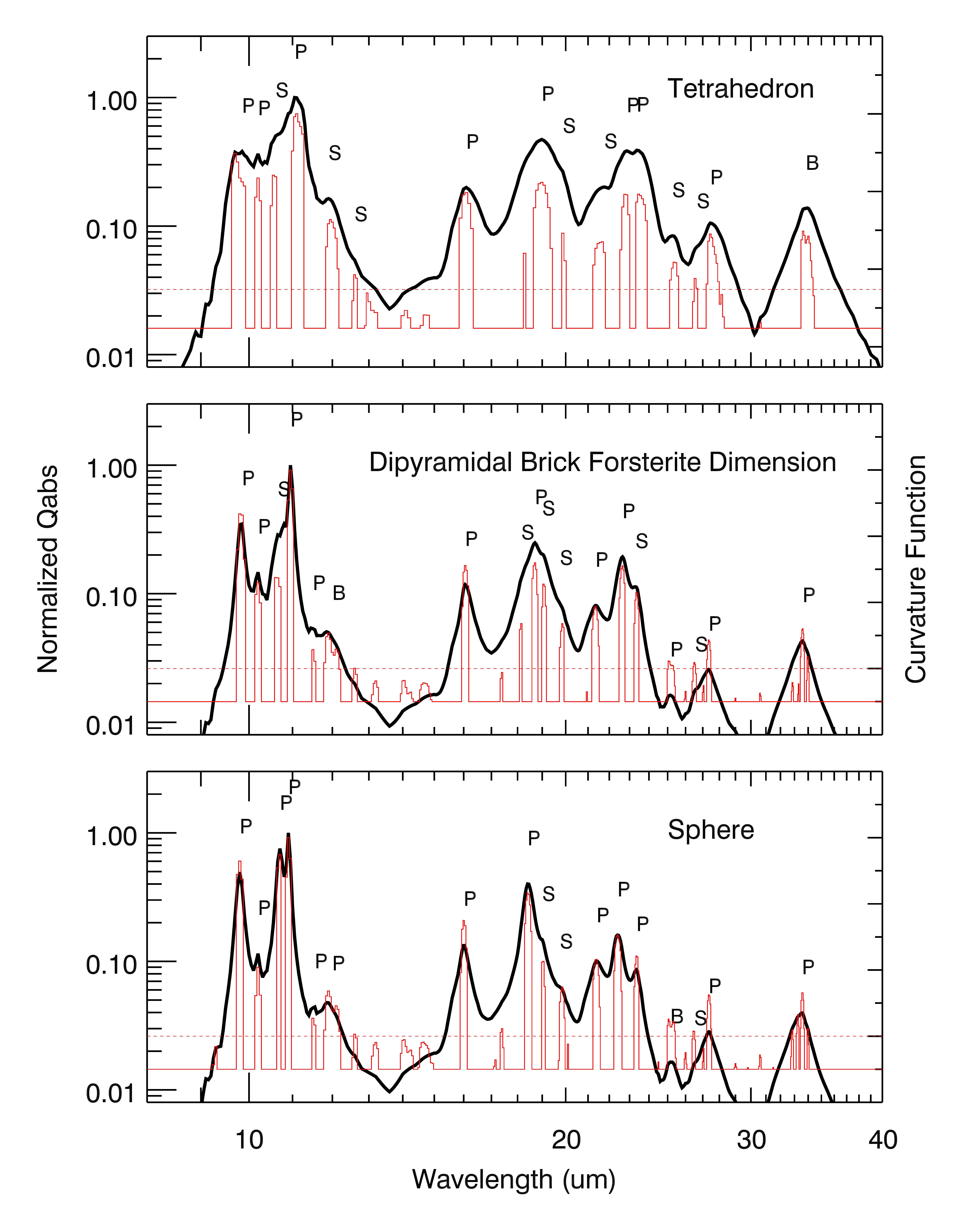,height=6.0in,width=5.0in}
        \caption{The curvature function (\emph{red}) compared to the DDSCAT $Q_{\rm abs}$ values (\emph{black}) for a tetrahedron (\emph{top}), dipyramidal brick with forsterite dimension (\emph{middle}), and sphere (\emph{bottom}) grain shapes.  The curvature function is defined as the square of the $Q_{\rm abs}$  divided by the smooth of the  $Q_{\rm abs}$ multiplied by the negative of the second derivative of the  $Q_{\rm abs}$.  The \emph{dotted red line} indicates the curvature strength cut-off value, such that features with curvature function value lower than the cut-off are not identified.  The letters above the features indicate the peaks (P), broad peaks (B),  and shoulders (S)}
        \label{fig:lo_method}
    \end{center}
\end{figure}

\end{document}